\documentclass[preprint,12pt, 3p]{elsarticle}
\usepackage{rotating}
\usepackage{graphicx}
\usepackage{booktabs}
\usepackage{amsmath}
\usepackage{amssymb}
\usepackage{adjustbox}
\usepackage{multirow}
\usepackage{subcaption}
\usepackage{xcolor}
\usepackage{float}

\newcommand{\BR}[1]{{#1}}

\definecolor{lightred}{RGB}{255,153,153}
\definecolor{darkred}{RGB}{139,0,0}
\definecolor{lightgreen}{RGB}{144,238,144}
\definecolor{darkgreen}{RGB}{0,100,0}
\definecolor{lightblue}{RGB}{135,206,250}
\definecolor{darkblue}{RGB}{0,0,139}
\definecolor{brown}{RGB}{165,42,42}

\newcommand{\VR}[1]{{#1}}

\begin{document}

\begin{frontmatter}
    \title{DReX: An Explainable Deep Learning-based Multimodal Recommendation Framework}    

    \author[1]{Adamya Shyam}
    \ead{ashyam@cs.du.ac.in}
    
    \author[2]{Venkateswara Rao Kagita}
    \ead{venkat.kagita@nitw.ac.in}
    
    \author[1]{Bharti Rana}
    \ead{bharti@cs.du.ac.in}

    \author[1]{Vikas Kumar\corref{cor1}}
    \ead{vikas@cs.du.ac.in}
    
    \address[1]{University of Delhi, Delhi, India}
    \address[2]{National Institute of Technology, Warangal, India}
    
    \cortext[cor1]{Corresponding author}

    \begin{abstract}
    Multimodal recommender systems leverage diverse data sources, such as user interactions, content features, and contextual information, to address challenges like cold-start and data sparsity. However, existing methods often suffer from one or more key limitations: processing different modalities in isolation, requiring complete multimodal data for each interaction during training, or independent learning of user and item representations. These factors contribute to increased complexity and potential misalignment between user and item embeddings. To address these challenges, we propose \textit{DReX}, a unified multimodal recommendation framework that incrementally refines user and item representations by leveraging interaction-level features from multimodal feedback. Our model employs gated recurrent units to selectively integrate these fine-grained features into global representations. This incremental update mechanism provides three key advantages: (1) simultaneous modeling of both nuanced interaction details and broader preference patterns, (2) eliminates the need for separate user and item feature extraction processes, leading to enhanced alignment in their learned representation, and (3) inherent robustness to varying or missing modalities. We evaluate the performance of the proposed approach on three real-world datasets containing reviews and ratings as interaction modalities. By considering review text as a modality, our approach automatically generates interpretable keyword profiles for both users and items, which supplement the recommendation process with interpretable preference indicators. Experiment results demonstrate that our approach outperforms state-of-the-art methods across all evaluated datasets.
    \end{abstract}
    
    \begin{keyword}
    Multimodal \sep%
    Explainability \sep%
    Recommender System \sep%
    Interaction-level Features \sep%
    Global Features
    \end{keyword}

\end{frontmatter}

\section{Introduction}
Recommender systems have become an integral part of digital platforms, helping users navigate vast amounts of data by providing personalized suggestions across domains like tourism~\cite{kbaier2017personalized}, entertainment~\cite{hwang2021movie} and healthcare~\cite{meng2022privacy}. Among various recommendation techniques, collaborative filtering (CF) has proven to be efficient in providing accurate and relevant recommendations to users while effectively addressing the challenge of information overload. Traditional CF approaches, such as matrix factorization~\cite{shyam2024unirecsys} and neighbourhood-based~\cite{valcarce2018finding} methods, utilize implicit or explicit feedback to generate personalized recommendations. These techniques operate on the principle that users who have exhibited similar behaviour in the past are likely to have similar interests in the future. Consequently, an item's recommendability is determined by analyzing historical interactions and identifying patterns of similarity among users or items. 

While effective in many scenarios, the traditional CF methods often struggle with data sparsity as they rely on a single modality of user-item interactions. Furthermore, models that use only ratings or clicks fail to capture the underlying reasoning behind user preferences. Multimodal recommender systems (MRS) address these challenges by extracting features from diverse data sources such as textual reviews, images, and ratings to capture a more accurate and interpretable understanding of user preferences and item characteristics~\cite{liu2024multimodal}. Moreover, these systems enhance explainability by aligning features extracted from different modalities with user preferences. For instance, textual reviews can provide explicit reasoning for recommendations by highlighting specific aspects that users appreciate. For example, if a user writes a review praising a movie’s “\textit{gripping storyline and complex characters},” the system can infer a preference for strong narratives and recommend similar films known for their compelling plots and well-developed characters. 

Several advancements have been proposed to enhance the performance of MRS.  McAuley et al.~\cite{mcauley2013hidden} introduced a statistical framework that integrates latent factors from rating data with topics extracted from review text. By integrating these two modalities, the approach improves interpretability by uncovering the natural alignment between rating dimensions and meaningful themes. Yang et al.~\cite{yang2018mmcf} proposed a multimodal CF-based model that enhances recommendations by integrating playlists and their diverse contents. The model comprises two key components: an autoencoder designed to capture the features of playlists and their associated artists and a character-level convolutional neural network that identifies latent connections between playlists and their titles. These components are trained independently, and their outputs are combined using an ensemble method during inference.
Liu et al.~\cite{liu2020hybrid} introduced a hybrid neural recommendation model that integrates both rating and review data. The model consists of a rating-based encoder to capture explicit user-item interaction patterns and a review-based encoder to extract latent features from textual reviews. The model makes use of an attention mechanism to identify and weigh important reviews for more accurate user and item modeling based on corresponding rating-based representations. 

Despite these advancements, several challenges continue to hinder the effective deployment and scalability of MRS. Many existing approaches treat different modalities independently, failing to leverage their complementary relationships. This independent processing of multimodal interaction prevents the system from fully capturing the intricate connections between user preferences and item characteristics. Additionally, some methods require complete multimodal data for every interaction due to the inherent joint modeling of multiple modalities in the objective function. This constraint is often impractical in real-world scenarios, where individual interactions frequently lack complete modality data, limiting the applicability of these approaches. Furthermore, several models train separate models for users and items, which not only increases computational complexity but also risks misalignment between the learned representations. As a result, these inefficiencies can lead to suboptimal recommendations and reduced interpretability.

In this work, we propose a novel neural-based explainable multimodal recommender framework, \textit{DReX}, to address the challenges with the traditional MRS. The proposed model focuses on learning a unified global representation for each user and item by exploiting the available multimodal interaction data. These global representations are iteratively refined using local features extracted from individual interactions, ensuring that the distinct attributes of users and items are accurately captured and integrated. Importantly, the local feature extraction process is designed to function without requiring all modality data for every interaction, making \textit{DReX} robust to missing or incomplete modalities. Additionally, since local features are extracted at the interaction level, the model is computationally efficient compared to approaches that train separate models for user and item feature extraction. To showcase the effectiveness of \textit{DReX}, we consider two modalities, namely reviews and ratings. 
First, the review contents are used to construct textual representations of users and items using an attention mechanism. Simultaneously, latent features are derived from the rating feedback. Thereafter, interaction-based local features are constructed by fusing textual representation and rating-based latent features. Finally, the local representations are selectively integrated into global user and item embeddings through gated recurrent units (GRUs), ensuring that the learned representations remain both dynamic and contextually relevant.

The inclusion of reviews as a modality highlights the model's ability to simultaneously generate user and item keyword profiles, providing interpretable preference indicators and improving predictive accuracy even when the rating information is limited.

\noindent The major contributions of the proposed framework can be summarized as follows.
\begin{itemize}
    \item \textit{Unified multimodal learning}: Integrates multiple modalities to learn unified user and item representations.
    \item \textit{Robust to missing modalities}: Unlike existing approaches that require complete data, \textit{DReX} handles missing modalities at the interaction level, making it suitable for real-world deployments.
    \item \textit{Interpretability}: Generates user and item keyword profiles to enhance recommendation explainability.
    \item \textit{Computational efficiency}: Eliminates separate feature extraction for learning user and item representations.
    \item \textit{Dynamic representation learning}: Uses GRUs to capture dynamic and context-aware user-item interactions.
    \item \textit{Empirical validation}:  Extensive experiments confirm the proposed framework's effectiveness in improving recommendation performance while ensuring explainability.
\end{itemize}

The remainder of this paper is structured as follows. Section \ref{relatedWork} discusses the existing related work on multimodal recommender systems. Section \ref{proposedDReX} introduces our proposed \textit{DReX} model. Section \ref{expSetup} presents detailed experimental results and analysis. Finally, Section \ref{concFuture} concludes the paper and explores future research directions.

\section{Related Works}
\label{relatedWork}
Collaborative filtering (CF) techniques utilize preferences of similar users along with the user's past preferences to provide recommendations~\cite{papadakis2022collaborative}. Matrix factorization (MF) is a common technique employed in many of the traditional CF models where the focus is to decompose the user-item interaction matrix into latent representations of users and items to provide similarity-based predictions~\cite{chen2022review}. Recent research has increasingly focused on neural network-based methods to model complex non-linear user-item relationships, achieving significant performance improvements in recommender systems~\cite{chen2020efficient}. Xue et al.\cite{xue2017deep} proposed a deep matrix factorization (DMF) method to perform recommendations using implicit feedback. DMF employs two parallel neural pipelines to learn the representations of users and items, which are then used to predict user preferences. Similarly, the DeepCoNN approach introduced by Zheng et al.~\cite{zheng2017joint} employs two parallel neural networks to model users and items based on available textual review data. Further, it utilizes a factorization machine to predict the ratings. While CF techniques have achieved significant accuracy in the prediction task, these techniques predominantly rely on a single modality of user-item interactions. This reliance inherently constraints their ability to generate accurate recommendations in scenarios with limited interaction feedback.

To mitigate the sparsity issue, multimodal recommendation systems (MRS) utilize multiple modalities, such as textual reviews, visual feedback, and/or implicit feedback, etc., for the effective modeling of user-item interaction. The core idea behind this approach is that different modalities capture distinct aspects of the entities. Leveraging these modalities jointly facilitates learning richer representations of user characteristics and item features~\cite{zhou2023comprehensive}. By utilizing the review text and implicit user-item feedback, Liu et al.~\cite{liu2019daml} proposed a dual attention mutual learning-based (DAML) approach. DAML learns an attention-based representation of review text through a convolutional neural network (CNN), which is then fused with rating features through a unified neural network. Finally, a factorization machine is used for the prediction. Building on the idea of DAML, HRDR model~\cite{liu2020hybrid} enhances representation learning through deeper integration of explicit ratings and review semantics. Two parallel pipelines with a review-level attention mechanism are used to extract useful reviews for users and items. The extracted reviews corresponding to a user or item are then aggregated to get the corresponding profile. He et al.~\cite{he2016vbpr} proposed VBPR, an extension of the Bayesian Personalized Ranking (BPR) framework that incorporates visual data to predict item rankings for users. VBPR learns \BR{visual representations of items} using a pretrained CNN model and integrates them with implicit feedback to capture variations in user preferences. Similarly, Chu et al.~\cite{chu2017hybrid} proposed a hybrid framework that incorporates visual information along with textual data to enhance restaurant recommendations. Several other works have also utilized multimodal interaction to enhance the recommendation accuracy and mitigate data sparsity issues~\cite{zhang2022latent, xiang2024neural}. 

Multiple studies have also focused on the idea of explainability to enhance the transparency of MRS~\cite{liu2023multimodal, chen2019personalized}. Chen et al.~\cite{chen2018neural} proposed a neural attention-based model (NARRE) that provides which reviews are useful in modeling a user or item profile as explanations. The model leverages an attention-based mechanism to learn the review-based features for users and items individually.  Alhejaili et al.~\cite{alhejaili2022expressive} employ weighted feature engineering on raw data to extract meaningful user and item features. These features are then passed to SVD for rating matrix reconstruction. This approach provides explanations based on the feature-engineered profiles and constructed rating matrix. In recent times, various studies have focused on generating natural language-based explanations for personalized recommendations~\cite{li2021personalized, wu2024pesi, li2023personalized}. Li et al.~\cite{li2021personalized} proposed a personalized transformer (PETER) to generate explanations along with predictions. PETER uses a context prediction task to bridge the semantic gap between user or item IDs and natural language, enabling personalized explanation generation.
In subsequent work, they introduced PEPLER, which utilizes the concept of prompt learning to generate explanations~\cite{li2023personalized}.
To address the problem of inconsistency between user ratings and review sentiments, Wu et al.~\cite{wu2024pesi} proposed PESI, a contrastive learning approach to separate shared and private features between ratings and reviews. The PESI approach ensures that the generated explanations account for the discrepancies in sentiments.

\section{Proposed Approach}
\label{proposedDReX}
In this section, we introduce \textit{DReX}, a novel multimodal recommender framework that aims at learning unified global representations for users and items. These global representations are iteratively refined through local representations extracted from individual multimodal interactions while maintaining robustness to missing modalities.  We have considered two modalities, reviews and ratings, to showcase the effectiveness of the proposed framework, while the model can accommodate any number of modalities. A key advantage of the \textit{DReX} model is its ability to handle textual interactions while providing interpretable explanations through learned user and item keyword profiles.

\begin{figure*}
    \centering
    \adjustbox{max width=\textwidth}{
    \includegraphics[scale = 1]{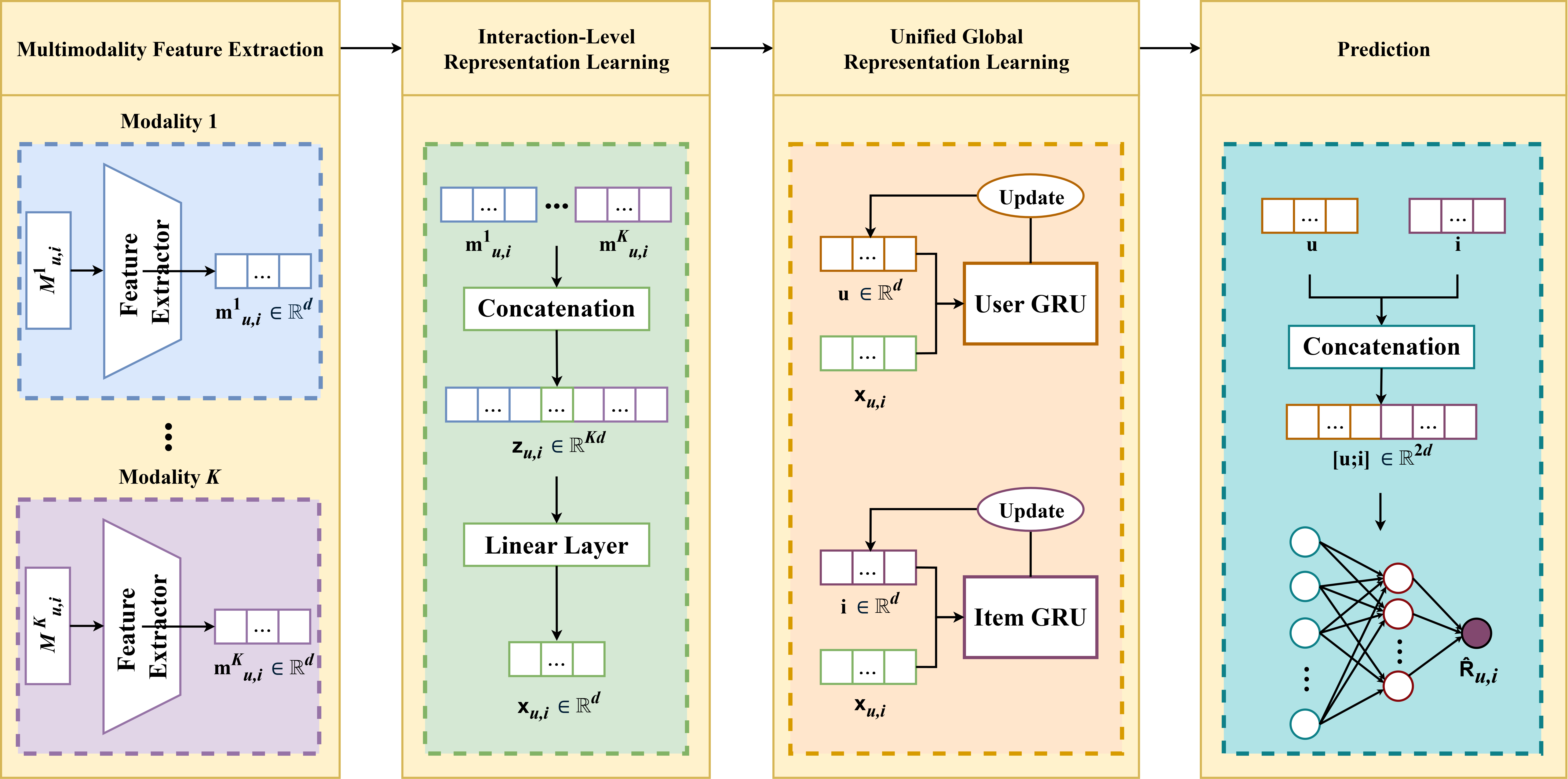}
    }
    \caption{Pipeline of the proposed DReX framework. All vectors are column-oriented.}
    \label{fig:pipeline}
\end{figure*}

The proposed framework consists of five key modules: (1) Modality-wise Feature Extraction, where features are independently extracted from each available modality within an interaction; (2) Interaction-Level Representation Learning, which integrates information from different modalities to construct interaction-level representations; (3) Unified Global Representation Learning, where the fused interaction-level features are used to refine the unified global representations of user and item;  (4) Rating Prediction, where the updated representations are used to generate personalized recommendations; and (5) Explanation, where learned keyword profiles of users and items are utilized to provide interpretable justifications for the recommendations. An outline of the proposed approach is shown in \VR{Figure \ref{fig:pipeline}}. 

In the following subsections, we first provide a formal definition of the problem and then present a detailed discussion of each module.

\subsection{Problem Formulation}
\label{probForm}
Let $\mathcal{U} = \{u_1, u_2, \dots, u_M\}$ denote a set of users and $\mathcal{I} = \{i_1, i_2, \dots, i_N\}$ denote a set of items. Each user-item interaction $(u,i)$ between user $u$ $(\in \mathcal{U})$ and item $i$ $ (\in \mathcal{I})$ is associated with multiple modalities, denoted as $\{M_{u,i}^1, M_{u,i}^2, \dots, M_{u,i}^K\}$, where $K$ is the total number of modalities. These modalities may include textual reviews, ratings, images, or other contextual information. \BR{The proposed} multimodal recommender system aims to leverage these diverse multimodal interactions to learn a function $f: \mathcal{U} \times \mathcal{I} \rightarrow \mathbb{R}$ that predicts the likelihood of any user $u$ interacting with an item $i$. 
Formally, the system aims to optimize: 
\begin{equation}
    f(u, i) = g(\mathbf{u}, \mathbf{i}, \phi(\psi_1(M_{u,i}^1), \dots, \psi_K(M_{u,i}^K))).  
\end{equation}

\noindent Here, $\mathbf{u} \in \mathbb{R}^d$ and $\mathbf{i} \in \mathbb{R}^d$ denote the unified global representations of the $u$th user and $i$th item, respectively, where $d$ is the embedding dimension.  $\psi_k(\cdot)$  is a feature extraction function for the $k$th modality, and $\phi(\cdot)$  is a fusion function that integrates information from the multimodal interaction associated with $(u, i)$. The function $g(\cdot)$  updates the unified global representations of user $u$  and item $i$  using the multimodal interaction level features obtained from $\psi_k(\cdot)$ and subsequently generates a recommendation score. The system is trained by minimizing a loss function that quantifies the discrepancy between the predicted and actual interactions, ensuring accurate and personalized recommendations.

\subsection{Modality-Specific Feature Extraction}
Feedback from each modality within a multimodal interaction $(u,i)$ is first processed through its respective feature extractor to generate interaction-level representations for both the user and item. These modality-specific features serve as the foundation for the subsequent fusion process, enabling a comprehensive understanding of user-item interactions. In this section, we describe the modality-specific feature extraction process, where we focus on two modalities, namely textual reviews and ratings. 

\begin{figure}[ht]
    \centering
    \includegraphics[width=0.55\linewidth]{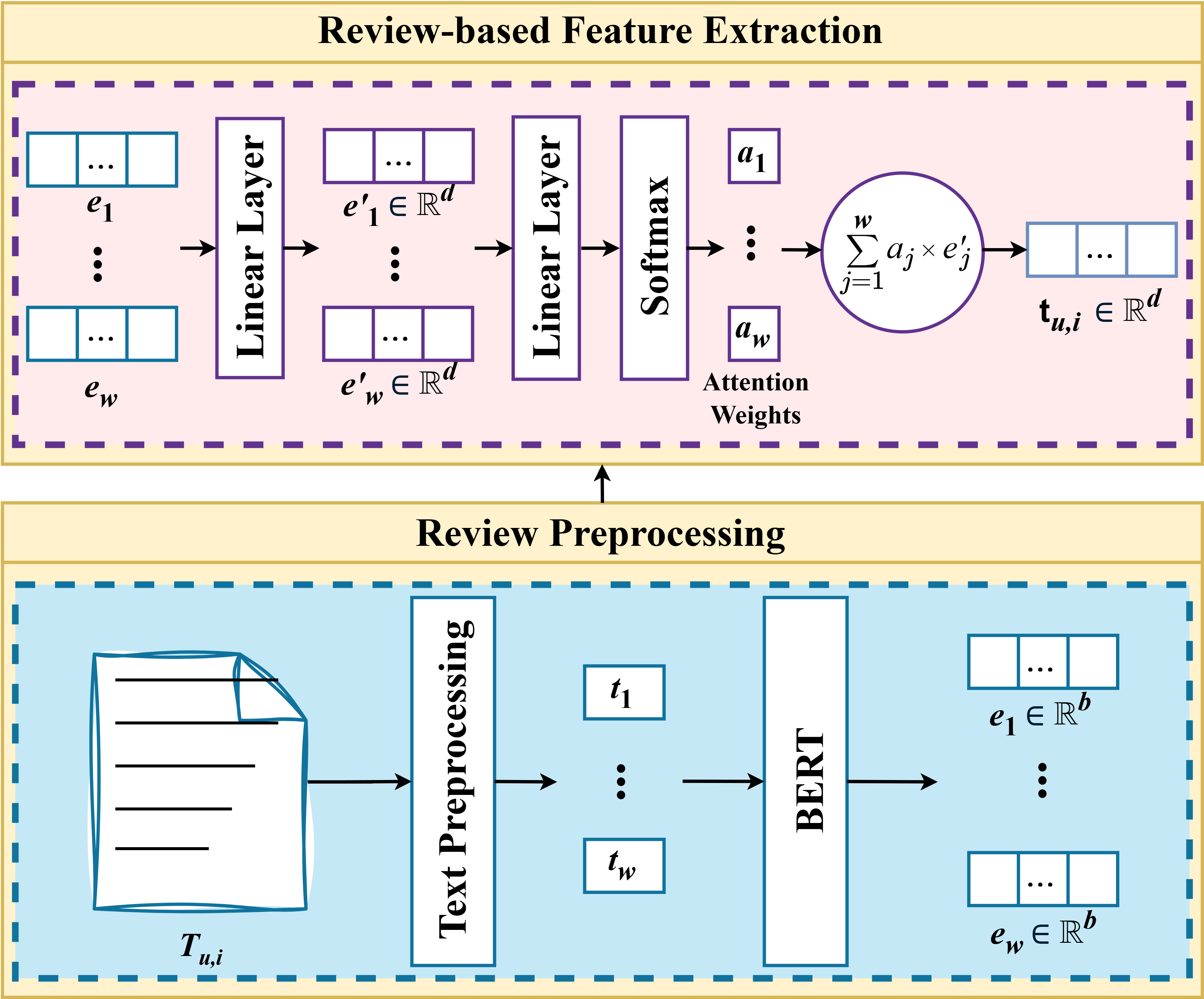}
    \caption{Outline of review-based feature extraction module.}
    \label{fig:revProcess}
\end{figure}

\subsubsection{Review-based Feature Extraction}
\label{revProcess}
The textual review associated with user-item interaction $(u, i)$, denoted as $T_{u, i}$, is passed through a multi-stage feature extraction process to capture meaningful semantic representations. First, the review text is preprocessed by removing stopwords, tokenizing, and applying lemmatization to ensure consistency in word forms, resulting in $w$ tokens.  Each token $t_j$, $j \in \{1,...,w\}$, is then encoded into a contextual embedding using Bidirectional Encoder Representations from Transformers (BERT) \cite{devlin2018bert}, producing a $b$-dimensional representation $\mathbf{e}_j$. 
Since BERT provides generic contextual embeddings, we apply a linear transformation to project each token embedding into a $d$-dimensional space more suitable for our recommendation task. We apply a linear transformation with a learnable weight matrix $P_t \in \mathbb{R}^{b \times d}$ and bias $\mathbf{b}_t \in \mathbb{R}^{d}$ to map each token embedding to a $ d $-dimensional space, yielding the transformed embeddings $\mathbf{e}'_j = P_t^\top \mathbf{e}_j + \mathbf{b}_t$.

To capture the importance of tokens in the review, we introduce an attention mechanism that assigns weights to each token based on its relevance. Specifically, we compute an alignment score for each token by applying a learnable linear transformation from the $d$-dimensional space to a scalar value, followed by a softmax function to obtain attention weights. Let $a_j \in \mathbb{R}$ denote the attention scores for the $j$th token, computed as:

\begin{equation}
   a_j = \frac{\exp(\mathbf{v}^\top \mathbf{e}'_j)}{\sum_{k=1}^{w} \exp(\mathbf{v}^\top \mathbf{e}'_k)} 
\end{equation}

\noindent where $\mathbf{v}$ is a learnable parameter vector. The final review representation is obtained through a weighted aggregation of token embeddings using these attention scores:  

\begin{equation}
\mathbf{t}_{u,i} = \sum_{j=1}^{w} a_j \mathbf{e}'_j
\end{equation}

This attention-weighted aggregation ensures that more informative tokens contribute more significantly to the final representation. Figure \ref{fig:revProcess} illustrates the feature extraction process from the review text.


\subsubsection{Rating-based Feature Extraction}

The numerical rating $R_{u,i} \in \{1, 2, \dots, S\}$ associated with each user-item interaction $(u,i)$ serves as an explicit indicator of user preferences across the rating scale. To incorporate this information into our model, we first convert the ordinal rating into a one-hot encoded vector  $\mathbf{s}_{u,i}$.  This sparse representation is then projected into a continuous $d$-dimensional latent space through a linear transformation as $\mathbf{r}_{u,i} = P_s^\top  \mathbf{s}_{u,i}+ \mathbf{b}_r$. Here, $P_s \in \mathbb{R}^{ S \times d}$ is a learnable weight matrix and $\mathbf{b}_r \in \mathbb{R}^{d}$ is a bias term. This transformation allows the model to capture the underlying relationships between different rating values and align them with other modalities. The rating-based representation $\mathbf{r}_{u,i}$ is then fused with textual features for a comprehensive user-item representation.

\subsection{Interaction-Level Representation Learning}

To model user-item interaction, we fuse modality-specific features into a unified representation.  Given the textual representation $ \mathbf{t}_{u,i} \in \mathbb{R}^d $ and the rating-based representation $ \mathbf{r}_{u,i} \in \mathbb{R}^d $, we construct an interaction-level feature vector $\mathbf{z}_{u,i} = [\mathbf{t}_{u,i} ; \mathbf{r}_{u,i}] \in \mathbb{R}^{2d}$.  
To refine these combined features into a more compact and meaningful form, we apply a learnable linear transformation with a learnable weight matrix $ P_x \in \mathbb{R}^{2d \times d} $ and bias $ \mathbf{b}_x \in \mathbb{R}^{d} $, mapping the concatenated feature vector into a $ d $-dimensional space as
$\mathbf{x}_{u,i} = P_x ^\top \mathbf{z}_{u,i} + \mathbf{b}_x$. This resulting interaction-level representation effectively captures both explicit and implicit user-item relationships, serving as the basis for updating global user and item embeddings.  \VR{\textit{DReX} can easily accommodate to missing interaction-level modalities by setting $\mathbf{z}_{u,i}[(k-1)d+1:kd]$, $1 \leq k \leq K$, to zero when the $k$th modality information is missing.}

\subsection{Unified Global Representation Learning} 
To capture the evolving preferences of users and dynamic characteristics of items, we update their unified global representations using a Gated Recurrent Unit (GRU)-based mechanism. Given the interaction-level embedding $\mathbf{x}_{u,i}$, we employ separate GRU networks for users and items to ensure that their representations are independently refined based on their respective interaction histories. This separation allows for the nuanced capture of both dynamic user preference shifts and the more gradual evolution of item popularity. While user preferences are often characterized by high variability, item characteristics exhibit greater stability, with changes primarily reflecting evolving user perceptions and valuations.

The GRU update equations for the $u$th user associated with interaction $(u,i)$ are given by: 
\begin{align}
\mathbf{r}_u &= \sigma(W_r^u \mathbf{u} + U_r^u \mathbf{x}_{u,i} + \mathbf{b}_r^u) \\
\mathbf{z}_u &= \sigma(W_z^u \mathbf{u} + U_z^u \mathbf{x}_{u,i} + \mathbf{b}_z^u) \\
\tilde{\mathbf{u}} ~&= \text{tanh}(W_h^u (\mathbf{r}_u \odot \mathbf{u}) + U_h^u \mathbf{x}_{u,i} + \mathbf{b}_h^u) \\
\mathbf{u}' &= (1 - \mathbf{z}_u) \odot \mathbf{u} + \mathbf{z}_u \odot \tilde{\mathbf{u}}
\end{align}

\noindent where $\mathbf{u}$ is the current global user representation, $\mathbf{r}_u$ is the reset gate, $\mathbf{z}_u$ is the update gate, and $\tilde{\mathbf{u}}$ is the candidate activation. The matrices ${W}_r^u, {W}_z^u, {W}_h^u \in \mathbb{R}^{d \times d}$ and $U_r^u, U_z^u, U_h^u \in \mathbb{R}^{d \times d}$ are learnable weight parameters, while $\mathbf{b}_r^u, \mathbf{b}_z^u, \mathbf{b}_h^u \in \mathbb{R}^d$ are bias terms. The symbol $\odot$ denotes element-wise multiplication, and $\sigma$ represents the sigmoid activation function. Similarly, $\mathbf{i}$, the item representation of $i$th item, is updated through a similar GRU mechanism with separate parameters.

\subsection{Rating Prediction}
\label{ratePred}
The learned global representations of user $u$ and item $i$ are first concatenated to form a joint representation, $[\mathbf{u};\mathbf{i}]$, which is then passed through a multi-layer perceptron (MLP) for rating prediction. We employ a multi-layer perceptron as the prediction layer with ReLU activation between layers to learn non-linear relationships between user and item representations. For our work, we have considered a two-layer perceptron with architecture $[2d \rightarrow d \rightarrow 1]$,  where the final output represents the predicted rating $\hat{R}_{u,i}$. We train the model end-to-end by minimizing a regularized mean squared error (MSE) loss:  
\begin{equation}
 \min_{U, I} J(U,I) = \sum_{(u,i) \in \Omega} (R_{u,i} - \hat{R}_{u,i})^2  + \frac{\lambda}{2}(\|\mathbf{u}\|_{F}^2+\|\mathbf{i}\|_{F}^2), 
\end{equation}
where $\mathbf{U} \in \mathbb{R}^{M \times d}$ is the user embedding matrix, where the $u$th row represents the global representation of user $u$, and $\mathbf{I} \in \mathbb{R}^{N \times d}$ is the item embedding matrix, where the $i$th row represents the global representation of item $i$. $\Omega$ represents the set of all interactions $(u, i)$ for which $R_{u,i}$ is available, $\hat{R}_{u,i}$ is the prediction corresponding to $R_{u,i}$, $\|.\|_F$ represents the Frobenius norm and $\lambda > 0$ is the trade-off parameter that balances the MSE loss and regularization. We employ the Adam optimizer with backpropagation to update all model parameters jointly, ensuring the learned representations adapt to both the collaborative filtering objective and the non-linear rating patterns. This design captures complex user-item relationships while maintaining model generalizability through explicit regularization. 

\subsection{Explanation}
\label{expPred}
In addition to minimizing the rating prediction error, the proposed approach uses the attention weight to simultaneously learn keyword profiles for users and items, which are later used to generate keyword-based explanations for users. Specifically, the $u$th user keyword profile $K_u$ is structured as a dictionary that stores all unique words appearing in her reviews $T_u$. The importance score of each word is determined by summing its attention weights across all occurrences while processing interactions involving user $u$ throughout the learning process, typically during the final iteration of training.  Item profiles $K_i$ are generated using the same approach. Finally, we retain the top-$k$ keywords based on their scores in the user and item profiles, ensuring that explanations remain concise and emphasize the most influential aspects of user preferences and item characteristics. The system appends the overlapping keywords between $K_u$ and $K_i$ as keyword-based explanations when recommending item $i$ to user $u$.

\section{Experiments}
\label{expSetup}
This section provides an overview of the datasets, baseline techniques, evaluation metrics, and hyperparameter tuning, followed by experimental results, including comparative and statistical analysis to validate the superiority of \textit{DReX}. In addition to reporting the results of \textit{DReX}, where the unified global representations of users and items are initialized randomly, we also present results for a variant, \textit{DReX-MLP}. Specifically, \textit{DReX-MLP} generates structured $d$-dimensional representations for each user and item using separate one-hidden-layer MLPs (hidden layer size $2d$), each with trainable parameters and ReLU activation functions. The user network takes user preference vectors of size $N$ as input, while the item network processes a vector of size $M$ constructed from the received preferences. 

\subsection{Datasets}
To evaluate the comparative performance of our proposed approach and the baselines, we utilize the publicly available benchmark datasets from Amazon\footnote{https://amazon-reviews-2023.github.io/} repository. We selected three diverse datasets from Video Games, Software, and CD \& Vinyl domains. 
The statistics of these datasets are presented in Table \ref{tab:datastats}.
\begin{table}[ht]
    \renewcommand{\arraystretch}{1}
    \centering
    \caption{Statistics of the selected datasets.}
    \adjustbox{max width=0.8\linewidth}{
        \begin{tabular}{lcccc}
        \hline
        Dataset & \# Users & \# Items & \# Interactions & Sparsity (\%) \\
        \hline
        Video Games  & 933      & 2347     & 31670           & 98.55  \\
        Software     & 3032     & 4027     & 78868           & 99.35  \\
        CD and Vinyl & 2260     & 5742     & 48704           & 99.62  \\       
        \hline
        \end{tabular}
        }
    \label{tab:datastats}
\end{table}
To reduce sparsity and enhance the reliability of learned representations, we filtered out users with fewer than twenty ratings and items rated by fewer than five users. The dataset was then split into training, testing, and validation sets in a $70:20:10$ ratio.

\subsection{Baselines}
We have considered five state-of-the-art methods for the comparative analysis and to demonstrate the superiority of the proposed approach.
Below, we provide a brief description of the selected baselines.  

\begin{itemize}
    \item EMF~\cite{abdollahi2016explainable}: Explainable Matrix Factorization (EMF) enhances the MF approach by integrating explainability into the recommendation process. The model leverages an explainability bipartite graph between users and items to quantify how explainable an item is to a user. An item is considered to be explainable to a user if a specific number of the user's neighbours have rated it. The model introduces an explainability regularization term in the loss function, guiding the latent factor learning toward more explainable recommendations.
    \item DMF~\cite{xue2017deep}: The Deep Matrix Factorization Model (DMF) extends the traditional MF approach by replacing linear projections with deep neural network-based nonlinear transformations.  This approach normalizes explicit ratings by dividing them by the maximum to convert them into implicit interactions. Thereafter, the user and item interaction vectors are passed through their corresponding multi-layer networks to learn their low-dimensional latent representations. The model makes use of a normalized cross-entropy loss function to optimize the learned representations and improve preference prediction.
    \item DeepCoNN~\cite{zheng2017joint}: The Deep Cooperative Neural Networks (DeepCoNN) model jointly learns user and item representations using only review information. The model consists of two parallel neural networks, one to learn user preferences by processing all reviews of the user, and the other to capture item characteristics from all reviews given to that item. Inspired by factorization machines, the parallel networks are coupled in the final layers through a shared layer. This allows the user and item latent features to interact, enhancing recommendation performance.
    \item NARRE~\cite{chen2018neural}: The Neural Attentional Regression with Review-level Explanation (NARRE) model argues that not all reviews are important for modeling user and item profiles. The model assigns attention weights to reviews based on their usefulness and filters out reviews with less weight to capture user preferences and item characteristics. The model learns the weights jointly with the recommendation task to maximize prediction accuracy. The model provides top reviews based on attention weights as an explanation for a prediction.
    \item PESI~\cite{wu2024pesi}: The Personalized Explanation with Sentiment Inconsistency (PESI) model addresses the misalignment between the rating and review sentiments to enhance the explainability. The model employs a novel disentanglement approach with contrastive learning to separate shared and private features between the two modalities. PESI consists of three integrated modules - rating prediction, explanation generation, and inconsistency extraction. The model uses entropy regularization to provide more coherent recommendations and generate explanations that accurately reflect users' true sentiments.
\end{itemize}

\begin{table}[!ht]
    \renewcommand{\arraystretch}{1}
    \centering
    \caption{Features of the comparing algorithms.}
    \adjustbox{max width=0.8\linewidth}{
        \begin{tabular}{lcccc}
        \hline
        \textbf{Model}                      & \textbf{Rating}   & \textbf{Review}   & \textbf{Explainable} & \textbf{Flexible}\\
        \hline
        EMF~\cite{abdollahi2016explainable} & $\checkmark$      &                   & $\checkmark$  & \\
        DMF~\cite{xue2017deep}              & $\checkmark$      &                   &            &    \\    
        DeepCoNN~\cite{zheng2017joint}      &                   & $\checkmark$      &              &  \\
        NARRE~\cite{chen2018neural}         & $\checkmark$      & $\checkmark$      & $\checkmark$  & \\
        PESI~\cite{wu2024pesi}              & $\checkmark$      & $\checkmark$      & $\checkmark$  & \\
        DReX (Proposed)   & $\checkmark$      & $\checkmark$      & $\checkmark$  & $\checkmark$ \\
        \hline
        \end{tabular}
    }
    \label{tab:baselines}
\end{table}

Though, multiple existing approaches incorporate ratings or reviews, they fail to provide an unified mechanism to fuse interaction-level multimodal information into joint user and item embeddings. Additionally, prior models lack the flexibility to operate in scenarios where some interactions have missing modality. under missing-modality scenarios. This distinguishing ability to dynamically refine shared representations establishes \textit{DReX} as a more flexible multimodal recommendation framework. Table \ref{tab:baselines} summarizes the characteristics of these models concerning interaction modalities and highlights the uniqueness of the proposed approach in comparison. 

\subsection{Evaluation Metrics}
We have assessed the performance of each comparative algorithm using three standard evaluation metrics: mean absolute error ($MAE$), $F1$-score, and normalized discounted cumulative gain ($NDCG$). Each of these metrics evaluates a different aspect of recommender system performance. 
Specifically, $MAE$ measures the accuracy of rating prediction by computing the absolute difference between actual and predicted ratings using the following formula:
\begin{equation}
    MAE = \frac{1}{|\Omega_t|} \sum \limits_{\{u,i\} \in \Omega_t}{|\hat{R}_{ui} - R_{ui}|}, 
\label{eq:mae}
\end{equation}
where $\Omega_t$ represents the set of user-item interactions in the test set for which ground truth ratings are available. A smaller value of $MAE$  indicates better rating prediction accuracy for the recommendation model.

The $F1$ score balances precision, which measures the proportion of relevant recommended items, and recall, which quantifies the proportion of relevant items that are recommended. $F1@k$ is the $F1$ score computed for the top-$k$ recommendations that balances Precision@k ($P@k$) and Recall@k ($R@k$) to evaluate the quality of the top-$k$ recommendations.
 
\begin{equation}
    F1@k = 2 \times \frac{P@k \times R@k}{P@k + R@k},
\label{eq:f1@k}
\end{equation}

\noindent where, 
\begin{equation}
P@k = \frac{1}{|\mathcal{U}_{t}|} \sum_{u \in \mathcal{U}_{t}} \frac{|rel_u(k)|}{k}, ~ R@k =  \frac{1}{|\mathcal{U}_{t}|} \sum_{u \in \mathcal{U}_{t}} \frac{|rel_u(k)|}{|rel_u|}.
\end{equation}
Here, $\mathcal{U}_{t}\subseteq \mathcal{U}$ set of users in the test set, $rel_u(k)$ is the set of relevant items for the $u^{th}$ user out of the top-$k$ suggestions, and $rel_u$ denotes the set of all relevant items for user $u$ in the test set. 
 
 $NDCG$ evaluates the ranking quality of the recommender model while considering the relevance and the order of recommendations. Mathematically, $NDCG$ is defined as follows:
\begin{equation}
NDCG@k = \frac{1}{|\mathcal{U}_{t}|} \sum_{u \in \mathcal{U}_{t}} \frac{\sum_{i=1}^k \frac{s_i^u}{log_2(i+1)}}{\sum_{i=1}^{|rel_u|} \frac{1}{log_2(i+1)}},
\label{eq:ndcg@k}
\end{equation}
where $s_i^u$ denotes the relevance score of the $i$th ranked item for user $u$. Both $F1@k$ and $NDCG@k$ scores are in the $[0,1]$ range, with $1$ denoting the best recommendation performance.

\begin{table}[ht]
    \centering
    \renewcommand{\arraystretch}{1}
    \caption{Optimal parameters for the comparing algorithms.}
    \adjustbox{max width=0.8\linewidth}{
    \begin{tabular}{lllllll}
    \hline
    \textbf{Dataset} & \textbf{Model}    & $\alpha$ & \textbf{Dropout} & $\lambda$  & $\beta$  & $\mathcal{G}$ \\
    \hline
    \multirow{7}{*}{Video Games}  
        & EMF~\cite{abdollahi2016explainable}      & 0.01          & -       & 0.0001 & 0.001 &     100 \\
        & DMF~\cite{xue2017deep}   & 0.0001        & -       & -      & -     & -   \\  
        & DeepCoNN~\cite{zheng2017joint} & 0.001   & 0       & -      & -     & -   \\
        & NARRE~\cite{chen2018neural}    & 0.1     & 0       & -      & -     & -   \\
        & PESI~\cite{wu2024pesi}   & 0.001         & 0.5     & 0.01   & -     & -   \\
        & DRex     & 0.1           & -       & 0.001  & -     & -   \\
        & DRex-MLP & 0.1           & -       & 0.0001 & -     & -   \\
    \hline
    \multirow{7}{*}{Software}     
        & EMF~\cite{abdollahi2016explainable}      & 0.01          & -       & 0.0001 & 0.001 & 20  \\
        & DMF~\cite{xue2017deep}      & 0.0001        & -       & -      & -     & -   \\
        & DeepCoNN~\cite{zheng2017joint} & 0.0001        & 0       & -      & -     & -   \\
        & NARRE~\cite{chen2018neural}    & 0.1           & 0.5     & 0      & -     & -   \\
        & PESI~\cite{wu2024pesi}     & 0.001         & 0       & 0.01   & -     & -   \\
        & DRex     & 0.1           & -       & 0.001  & -     & -   \\
        & DRex-MLP & 0.01          & -       & 0.001  & -     & -   \\
    \hline
    \multirow{7}{*}{CD and Vinyl} 
        & EMF~\cite{abdollahi2016explainable}      & 0.01          & -       & 0.0001 & 0.01  & 20  \\
        & DMF~\cite{xue2017deep}      & 0.0001        & -       & -      & -     & -   \\
        & DeepCoNN~\cite{zheng2017joint} & 0.001         & 0       & 0      & -     & -   \\
        & NARRE~\cite{chen2018neural}    & 0.001         & 0       & 0      & -     & -   \\
        & PESI~\cite{wu2024pesi}     & 0.001         & 0.1     & 0.01   & -     & -   \\
        & DRex     & 0.1           & -       & 0.0001 & -     & -   \\
        & DRex-MLP & 0.1           & -       & 0.0001 & -     & -   \\
    \hline
    \end{tabular}}
    \label{tab:optset}
\end{table}

\subsection{Parameter Configuration}
For a fair comparison, we performed hyperparameter tuning for all algorithms. The learning rate, $\alpha$ is tuned in $[1e-4, 1e-3, 1e-2, 1e-1]$. The regularization parameter, $\lambda$ and the final embedding dimension, $d$ are tuned in $[0.0001, 0.001, 0.01, 0.1, 1, 10, 100]$, and $[8, 16, 32, 64, 128, 256]$, respectively. We use the Adam optimizer with a batch size of $1024$ and a maximum of 100 epochs. For $EMF$, we have tuned the number of groups, $\mathcal{G}$ in $[10, 20, 30, \cdots, 90, 100]$ and the explainability coefficient $\beta$ in $[0.0001, 0.001, 0.01, 0.1, 1, 10, 100]$. For $DeepCoNN$ and $NARRE$, we have considered dropout in $[0.0, 0.1, 0.3, 0.5, 0.7, 0.9]$. The convolutional layer size and window size are fixed at $100$ and $3$, respectively, for $DeepCoNN$ and $NARRE$, and Google News\footnote{https://code.google.com/archive/p/word2vec/} is used to obtain the pre-trained word embeddings, as reported in \cite{chen2018neural}. For \textit{DReX}, we have utilized the \textit{bert-base-uncased}\footnote{https://huggingface.co/google-bert/bert-base-uncased} model with $b=768$ to obtain the contextual embedding of each token.  To avoid overfitting, early stopping is applied based on the average $F1@5$ and $NDCG@5$ on the validation set with $patience = 10$.

\begin{figure*}[!ht]
    \centering        
        \begin{subfigure}[b]{0.32\textwidth}
            \centering
            \includegraphics[height  = 5 cm ]{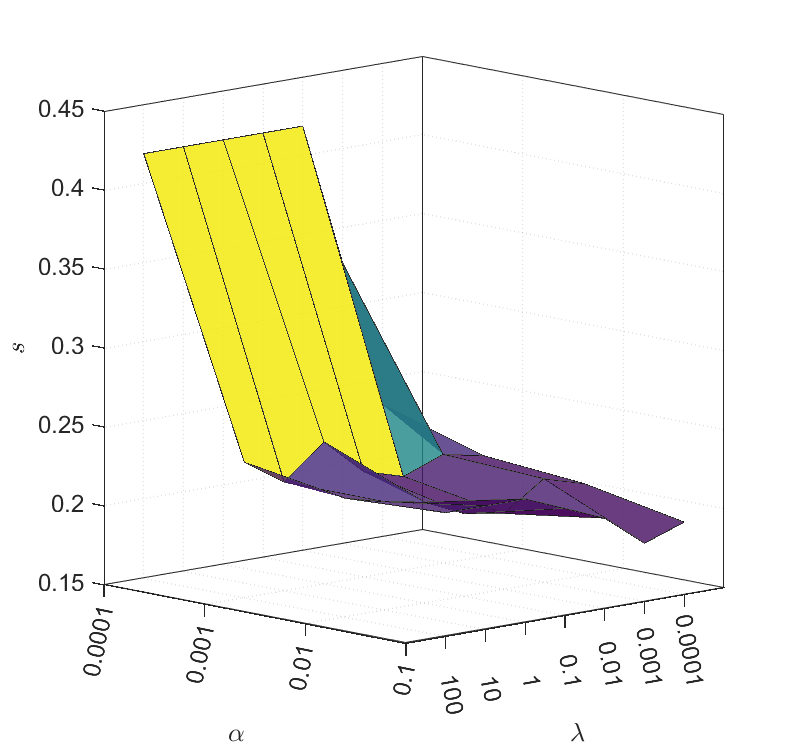}
            \caption{Video Games}
            \label{fig:drex_vg}
        \end{subfigure}
         \hfill
        \begin{subfigure}[b]{0.32\textwidth}
            \centering
            \includegraphics[height  = 5 cm]{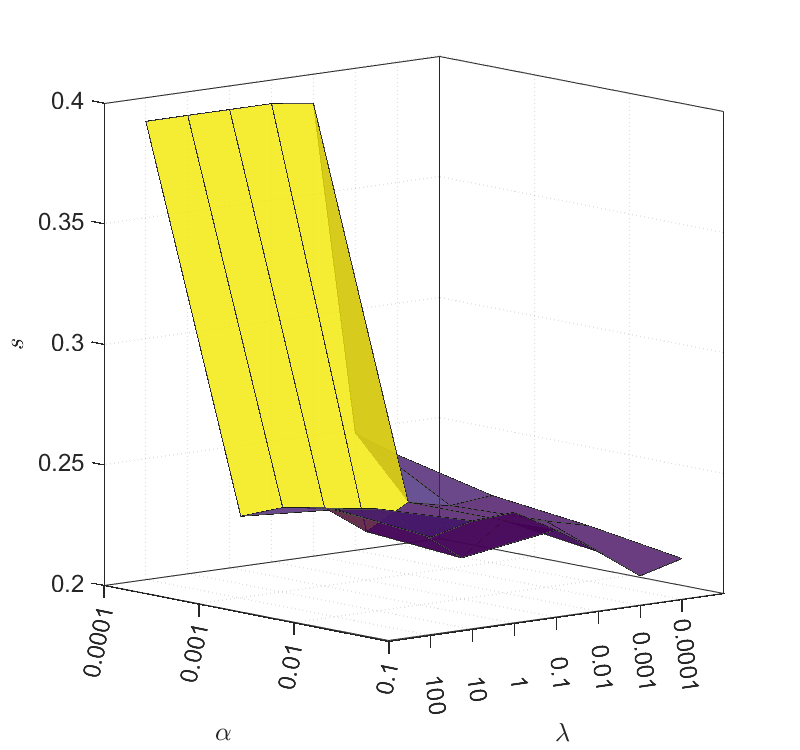}
            \caption{Software}
            \label{fig:drex_sw}
        \end{subfigure}
         \hfill
        \begin{subfigure}[b]{0.32\textwidth}
            \centering
            \includegraphics[height  = 5 cm]{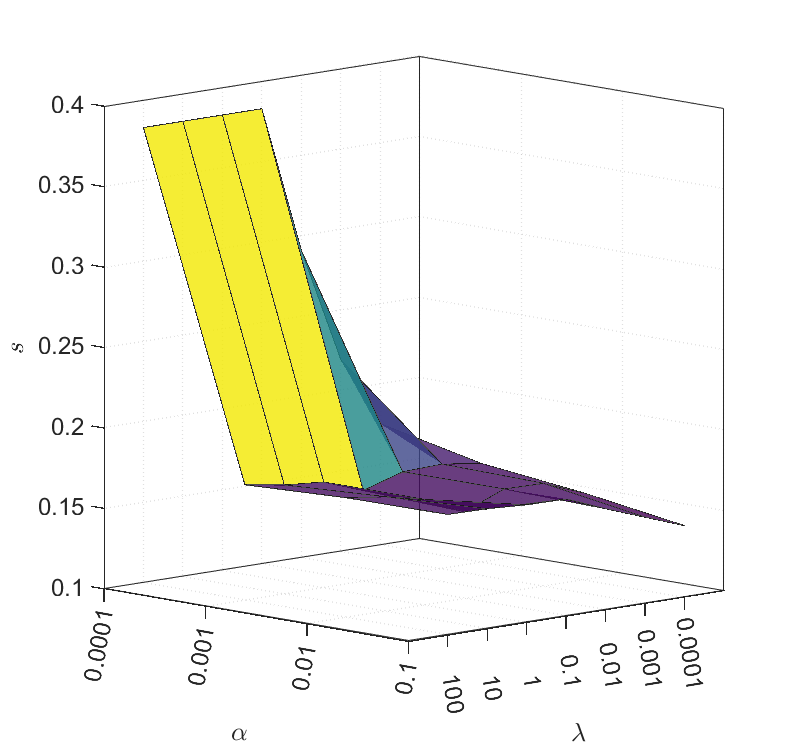}
            \caption{CD and Vinyl}
            \label{fig:drex_cdnv}
        \end{subfigure}
        
    \caption{Impact of learning rate ($\alpha$) and regularization ($\lambda$) on the DReX approach.}
    \label{fig:mesh_lr_lambda}
\end{figure*}

Furthermore, to select the best set of hyperparameters we computed a \textit{score} $s$ as the average across three runs of $MAE$, $1-F1@k$, and $1-NDCG@k$ for $k = \{1,2,3,4,5\}$.  For a fair comparison, we fixed the final latent space embedding size at 64 for all models. Figure \ref{fig:mesh_lr_lambda} depicts the influence of learning rate, $\alpha$, and the regularization $\lambda$ on our proposed models over the three datasets. It can be seen that the proposed \textit{DReX} models perform better for higher learning rates and lower regularization values over each of the datasets. For $d = 64$, the optimal hyperparameter set for each comparing algorithm is reported in Table \ref{tab:optset}. All results reported in the subsequent section are averaged over three runs on the test set.

\begin{figure*}[h]
    \centering
        
        \begin{subfigure}[b]{0.32\textwidth}
            \centering
            \includegraphics[width=\textwidth]{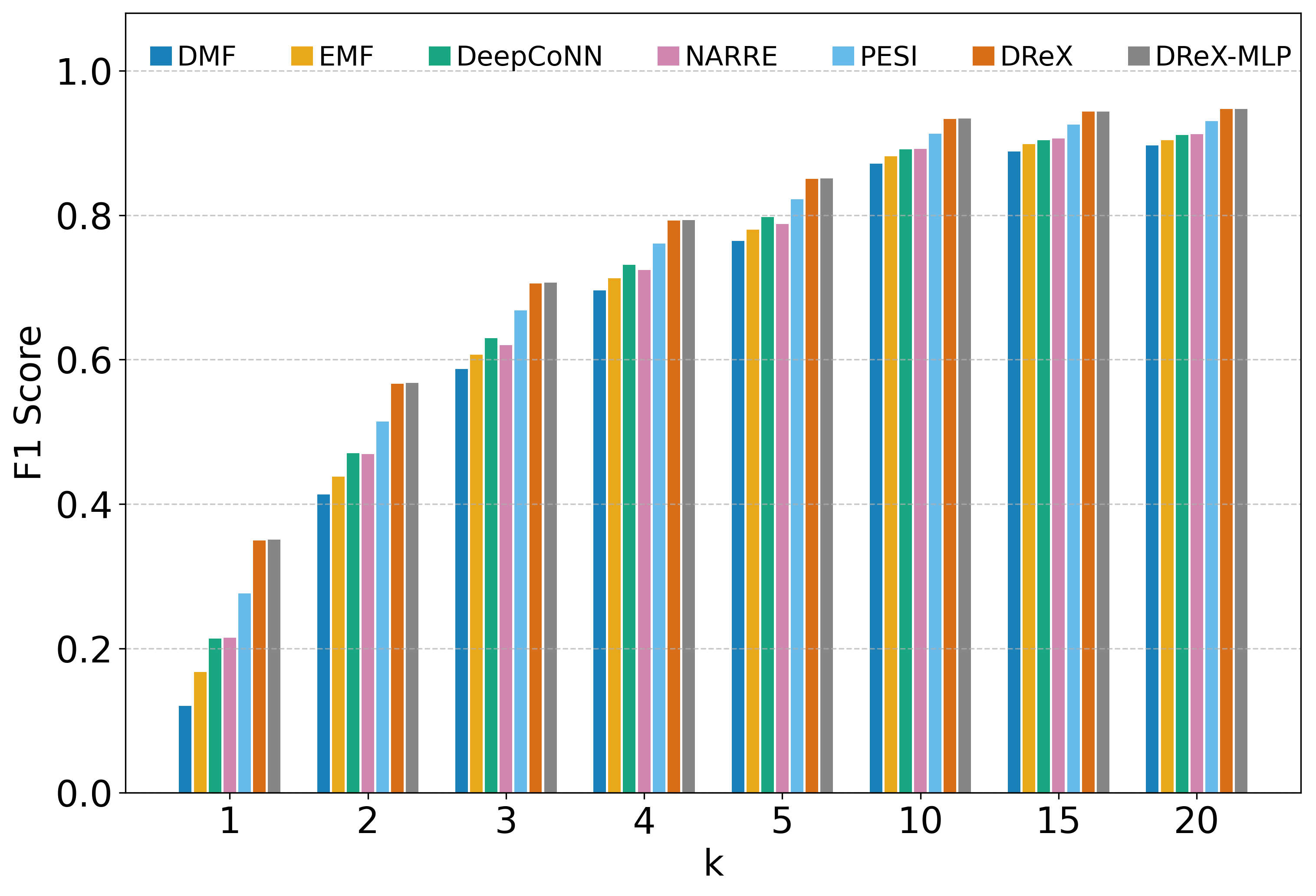}
            \caption{Video Games}
            \label{fig:f1_vg}
        \end{subfigure}
        \begin{subfigure}[b]{0.32\textwidth}
            \centering
            \includegraphics[width=\textwidth]{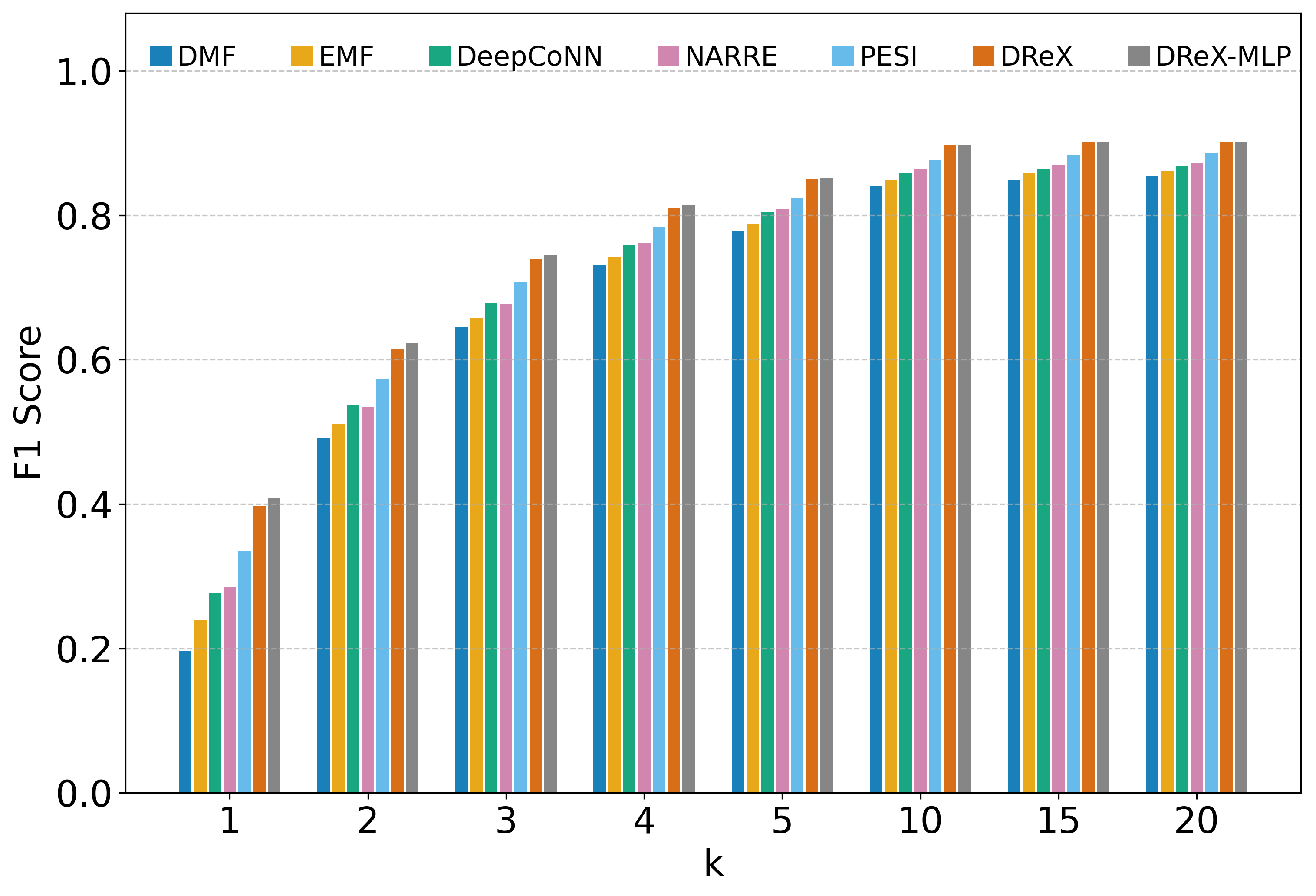}
            \caption{Software}
            \label{fig:f1_sw}
        \end{subfigure}
        \begin{subfigure}[b]{0.32\textwidth}
            \centering
            \includegraphics[width=\textwidth]{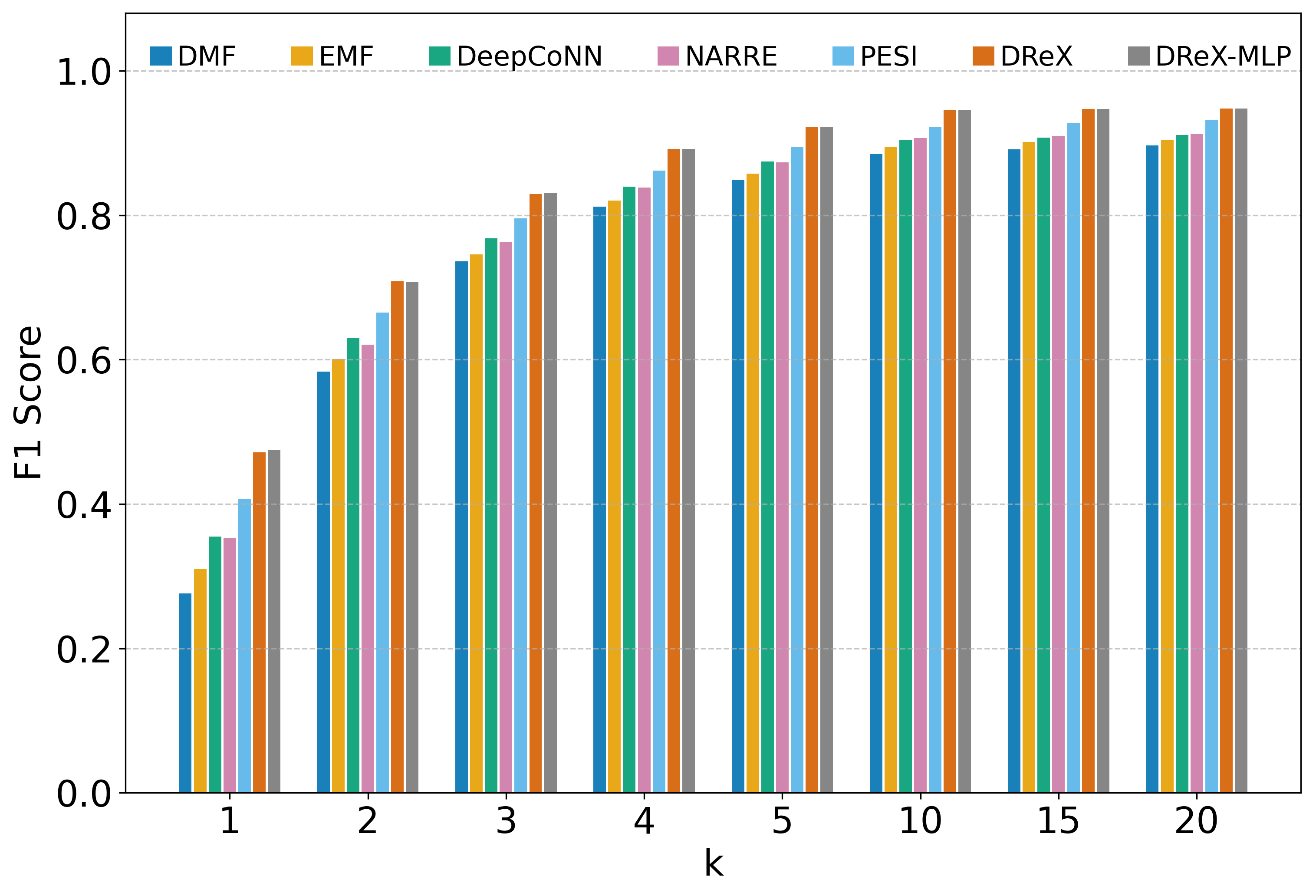}
            \caption{CD and Vinyl}
            \label{fig:f1_cdnv}
        \end{subfigure}
        
    \caption{Performance of comparing algorithms in terms of \textit{F1-Score}.}
    \label{fig:f1_scores}
\end{figure*}

\subsection{Empirical Analysis} 
Table \ref{tab:results} illustrates the comparative performance of each algorithm based on $MAE$ and $NDCG@k$ for $k\in \{1,2,3,4,5,10,15,20\}$ across all three datasets. The results indicate that the proposed variants achieve superior performance in terms of $MAE$ on two datasets while exhibiting a slightly higher $MAE$ for the Software dataset. Additionally, the proposed variants significantly outperform all baselines in terms of $NDCG$, highlighting its superior ranking capabilities. 

\renewcommand{\arraystretch}{1.8}
\begin{sidewaystable*}[]
\caption{Results of the comparing algorithms (mean$\pm$standard deviation rank) in terms of $MAE$ and $NDCG@k$. (\textit{$\uparrow$" indicates that a higher value is better, while $\downarrow$" indicates that a lower value is better)}}
\adjustbox{max width = \linewidth}{
\begin{tabular}{llllllllllllllllllll}
\hline
\textbf{Dataset}                       & \textbf{Model}    & \textbf{MAE} ($\downarrow$)   &          & \textbf{NDCG@1} ($\uparrow$) &         & \textbf{NDCG@2} ($\uparrow$)  &        & \textbf{NDCG@3} ($\uparrow$)  &         & \textbf{NDCG@4} ($\uparrow$)    &      & \textbf{NDCG@5} ($\uparrow$)  &        & \textbf{NDCG@10} ($\uparrow$)  &       & \textbf{NDCG@15} ($\uparrow$)  &       & \textbf{NDCG@20} ($\uparrow$) &        \\ \hline
\multirow{7}{*}{\textbf{Video Games}}  
    & \textbf{EMF}~\cite{abdollahi2016explainable}      & 0.7998 $\pm$ 0.0343          & 5          & 0.4197 $\pm$ 0.0014          & 6          & 0.5372 $\pm$ 0.0027          & 6          & 0.6104 $\pm$ 0.0029          & 6          & 0.6603 $\pm$ 0.0029          & 6          & 0.6978 $\pm$ 0.0033          & 6          & 0.7596 $\pm$ 0.0021          & 6          & 0.7825 $\pm$ 0.0013          & 6          & 0.7908 $\pm$ 0.0013          & 6          \\
    & \textbf{DMF}~\cite{xue2017deep}      & -                        & -          & 0.3752 $\pm$ 0.0054          & 7          & 0.4935 $\pm$ 0.0055          & 7          & 0.5740 $\pm$ 0.0042           & 7          & 0.6337 $\pm$ 0.0038          & 7          & 0.6685 $\pm$ 0.0012          & 7          & 0.7403 $\pm$ 0.0019          & 7          & 0.7575 $\pm$ 0.0012          & 7          & 0.7646 $\pm$ 0.0010           & 7          \\
    & \textbf{DeepCoNN}~\cite{zheng2017joint} & 0.6673 $\pm$ 0.0089          & 2          & 0.4864 $\pm$ 0.0161          & 5          & 0.5962 $\pm$ 0.0115          & 4          & 0.6628 $\pm$ 0.0076          & 4          & 0.7090 $\pm$ 0.0047           & 4          & 0.7392 $\pm$ 0.0043          & 4          & 0.793 $\pm$ 0.0020            & 4          & 0.8082 $\pm$ 0.0014          & 5          & 0.8202 $\pm$ 0.0011          & 5          \\
    & \textbf{NARRE}~\cite{chen2018neural}    & 1.4325 $\pm$ 0.3562          & 6          & 0.4899 $\pm$ 0.0150           & 4          & 0.5904 $\pm$ 0.0146          & 5          & 0.6580 $\pm$ 0.0142           & 5          & 0.7061 $\pm$ 0.0139          & 5          & 0.7358 $\pm$ 0.0135          & 5          & 0.7901 $\pm$ 0.0121          & 5          & 0.8104 $\pm$ 0.0102          & 4          & 0.8282 $\pm$ 0.0078          & 4          \\
    & \textbf{PESI}~\cite{wu2024pesi}    & 0.6836 $\pm$ 0.0460           & 4          & 0.6104 $\pm$ 0.0388          & 3          & 0.7120 $\pm$ 0.0275           & 3          & 0.7774 $\pm$ 0.0186          & 3          & 0.8215 $\pm$ 0.0112          & 3          & 0.8511 $\pm$ 0.0073          & 3          & 0.8990 $\pm$ 0.0015           & 3          & 0.9156 $\pm$ 0.0011          & 3          & 0.9200 $\pm$ 0.0009            & 3          \\
    & \textbf{DReX}     & \textbf{0.6531 $\pm$ 0.016}  & \textbf{1} & \textbf{0.6720 $\pm$ 0.0114}  & \textbf{1} & \textbf{0.7718 $\pm$ 0.0094} & \textbf{1} & \textbf{0.8367 $\pm$ 0.0084} & \textbf{1} & \textbf{0.8815 $\pm$ 0.0040}  & \textbf{1} & \textbf{0.9118 $\pm$ 0.0025} & \textbf{1} & 0.9578 $\pm$ 0.0010           & 2          & 0.9645 $\pm$ 0.0003          & 2          & 0.9663 $\pm$ 0.0002          & 2          \\
    & \textbf{DReX-MLP}   & 0.6677 $\pm$ 0.0141          & 3          & 0.6490 $\pm$ 0.0119           & 2          & 0.7616 $\pm$ 0.0063          & 2          & 0.8324 $\pm$ 0.0041          & 2          & 0.8785 $\pm$ 0.0045          & 2          & 0.9109 $\pm$ 0.0029          & 2          & \textbf{0.9595 $\pm$ 0.0011} & \textbf{1} & \textbf{0.9661 $\pm$ 0.0013} & \textbf{1} & \textbf{0.9680 $\pm$ 0.0014}  & \textbf{1} \\ \hline
\multirow{7}{*}{\textbf{Software}}     
    & \textbf{EMF}~\cite{abdollahi2016explainable}      & 0.9770 $\pm$ 0.0107           & 5          & 0.4803 $\pm$ 0.0015          & 6          & 0.5927 $\pm$ 0.0033          & 6          & 0.6528 $\pm$ 0.0036          & 6          & 0.6911 $\pm$ 0.0032          & 6          & 0.7164 $\pm$ 0.0029          & 6          & 0.7538 $\pm$ 0.0022          & 6          & 0.7726 $\pm$ 0.0020           & 6          & 0.7794 $\pm$ 0.0020           & 6          \\
    & \textbf{DMF}~\cite{xue2017deep}      & -                        & -          & 0.4227 $\pm$ 0.0072          & 7          & 0.5408 $\pm$ 0.0041          & 7          & 0.6139 $\pm$ 0.0006          & 7          & 0.6635 $\pm$ 0.0015          & 7          & 0.6890 $\pm$ 0.0012           & 7          & 0.7380 $\pm$ 0.0011           & 7          & 0.7500 $\pm$ 0.0008            & 7          & 0.7557 $\pm$ 0.0006          & 7          \\
    & \textbf{DeepCoNN}~\cite{zheng2017joint} & 0.8292 $\pm$ 0.0412          & 4          & 0.5162 $\pm$ 0.0245          & 4          & 0.6189 $\pm$ 0.0144          & 4          & 0.6822 $\pm$ 0.0082          & 5          & 0.7237 $\pm$ 0.0044          & 5          & 0.7480 $\pm$ 0.0035           & 5          & 0.7866 $\pm$ 0.001           & 5          & 0.7977 $\pm$ 0.0010           & 5          & 0.8085 $\pm$ 0.0011          & 5          \\
    & \textbf{NARRE}~\cite{chen2018neural}     & 1.3188 $\pm$ 0.1099          & 6          & 0.5023 $\pm$ 0.0086          & 5          & 0.6154 $\pm$ 0.0085          & 5          & 0.6938 $\pm$ 0.0084          & 4          & 0.7357 $\pm$ 0.0083          & 4          & 0.7602 $\pm$ 0.0082          & 4          & 0.7940 $\pm$ 0.0075           & 4          & 0.8086 $\pm$ 0.0067          & 4          & 0.8245 $\pm$ 0.0058          & 4          \\
    & \textbf{PESI}~\cite{wu2024pesi}      & \textbf{0.7723 $\pm$ 0.0145} & \textbf{1} & 0.6666 $\pm$ 0.0155          & 2          & 0.7578 $\pm$ 0.0094          & 3          & 0.8096 $\pm$ 0.0063          & 3          & 0.8413 $\pm$ 0.0043          & 3          & 0.8604 $\pm$ 0.0032          & 3          & 0.8974 $\pm$ 0.0011          & 3          & 0.9038 $\pm$ 0.0009          & 3          & 0.9100 $\pm$ 0.0008            & 3          \\
    & \textbf{DReX}     & 0.8151 $\pm$ 0.0203          & 3          & 0.6512 $\pm$ 0.0053          & 3          & 0.7706 $\pm$ 0.0019          & 2          & 0.8438 $\pm$ 0.0014          & 2          & 0.8911 $\pm$ 0.0021          & 2          & 0.9185 $\pm$ 0.0020           & 2          & 0.9524 $\pm$ 0.0009          & 2          & 0.9552 $\pm$ 0.0007          & 2          & 0.9557 $\pm$ 0.0006          & 2          \\
    & \textbf{DReX-MLP}   & 0.7977 $\pm$ 0.0208          & 2          & \textbf{0.6765 $\pm$ 0.0098} & \textbf{1} & \textbf{0.7904 $\pm$ 0.0065} & \textbf{1} & \textbf{0.8579 $\pm$ 0.0062} & \textbf{1} & \textbf{0.9017 $\pm$ 0.0045} & \textbf{1} & \textbf{0.9264 $\pm$ 0.0040}  & \textbf{1} & \textbf{0.9568 $\pm$ 0.0022} & \textbf{1} & \textbf{0.9592 $\pm$ 0.0019} & \textbf{1} & \textbf{0.9596 $\pm$ 0.0017} & \textbf{1} \\
 \hline
\multirow{7}{*}{\textbf{CD and Vinyl}} 
    & \textbf{EMF}~\cite{abdollahi2016explainable}      & 0.9113 $\pm$ 0.0230           & 5          & 0.4896 $\pm$ 0.0040           & 6          & 0.6110 $\pm$ 0.0009           & 6          & 0.6770 $\pm$ 0.0011           & 6          & 0.7170 $\pm$ 0.0007           & 6          & 0.7397 $\pm$ 0.0009          & 6          & 0.7689 $\pm$ 0.0007          & 6          & 0.7867 $\pm$ 0.0008          & 6          & 0.7934 $\pm$ 0.0008          & 6          \\
    & \textbf{DMF}~\cite{xue2017deep}      & -                        & -          & 0.4484 $\pm$ 0.0044          & 7          & 0.573 $\pm$ 0.0018           & 7          & 0.6487 $\pm$ 0.0008          & 7          & 0.6965 $\pm$ 0.0009          & 7          & 0.7172 $\pm$ 0.0010           & 7          & 0.7540 $\pm$ 0.0006           & 7          & 0.7648 $\pm$ 0.0006          & 7          & 0.7704 $\pm$ 0.0006          & 7          \\
    & \textbf{DeepCoNN}~\cite{zheng2017joint} & 0.6968 $\pm$ 0.0180           & 4          & 0.5345 $\pm$ 0.0141          & 5          & 0.6511 $\pm$ 0.0070           & 5          & 0.7208 $\pm$ 0.0028          & 5          & 0.7595 $\pm$ 0.0004          & 5          & 0.7788 $\pm$ 0.0008          & 5          & 0.8052 $\pm$ 0.0003          & 5          & 0.8151 $\pm$ 0.0003          & 5          & 0.8257 $\pm$ 0.0004          & 5          \\
    & \textbf{NARRE}~\cite{chen2018neural}     & 1.6496 $\pm$ 0.1340           & 6          & 0.5534 $\pm$ 0.0085          & 4          & 0.6701 $\pm$ 0.0083          & 4          & 0.7353 $\pm$ 0.0081          & 4          & 0.7699 $\pm$ 0.0081          & 4          & 0.7877 $\pm$ 0.0079          & 4          & 0.8081 $\pm$ 0.0075          & 4          & 0.8218 $\pm$ 0.0064          & 4          & 0.8370 $\pm$ 0.0058           & 4          \\
    & \textbf{PESI}~\cite{wu2024pesi}     & 0.6800 $\pm$ 0.0345            & 2          & 0.6482 $\pm$ 0.0191          & 3          & 0.7656 $\pm$ 0.0100            & 3          & 0.8400 $\pm$ 0.0065            & 3          & 0.8745 $\pm$ 0.0036          & 3          & 0.8920 $\pm$ 0.0024           & 3          & 0.9100 $\pm$ 0.0011            & 3          & 0.9145 $\pm$ 0.0010           & 3          & 0.9257 $\pm$ 0.0010          & 3          \\
    & \textbf{DReX}     & \textbf{0.6441 $\pm$ 0.0177} & \textbf{1} & \textbf{0.7333 $\pm$ 0.0234} & \textbf{1} & \textbf{0.8455 $\pm$ 0.0114} & \textbf{1} & \textbf{0.9099 $\pm$ 0.0049} & \textbf{1} & \textbf{0.9430 $\pm$ 0.0037}  & \textbf{1} & \textbf{0.9594 $\pm$ 0.0025} & \textbf{1} & 0.9732 $\pm$ 0.0013          & 2          & 0.9743 $\pm$ 0.0012          & 2          & 0.9746 $\pm$ 0.0013          & 2          \\
    & \textbf{DReX-MLP}   & 0.6850 $\pm$ 0.0445           & 3          & 0.7073 $\pm$ 0.0244          & 2          & 0.8305 $\pm$ 0.0137          & 2          & 0.9019 $\pm$ 0.0084          & 2          & 0.9402 $\pm$ 0.0047          & 2          & 0.9591 $\pm$ 0.0030           & 2          & \textbf{0.9742 $\pm$ 1.0011} & \textbf{1} & \textbf{0.9751 $\pm$ 0.0009} & \textbf{1} & \textbf{0.9754 $\pm$ 0.0009} & \textbf{1}
 \\ \hline
\end{tabular}}

\label{tab:results}
\end{sidewaystable*}

Comparing the two variants, \textit{DReX-MLP} outperforms \textit{DReX} in $NDCG@k$ for larger $k$. This can be attributed to long-tail recommendation dynamics, where \textit{DReX} optimizes local item relevance for smaller $k$, while \textit{DReX-MLP}, with its additional MLPs, captures global user-item relationships, making it more effective for larger $k$ where long-tail items become more significant. The results for the $F1$ score are reported in Figure \ref{fig:f1_scores}. The results reveal that the proposed variants significantly outperform the state-of-the-art methods by a significant margin. This underscores that our framework not only achieves better accuracy but also delivers more contextually relevant items to users. 

\begin{figure}[!h]
    \centering
        \includegraphics[width=0.4\textwidth]{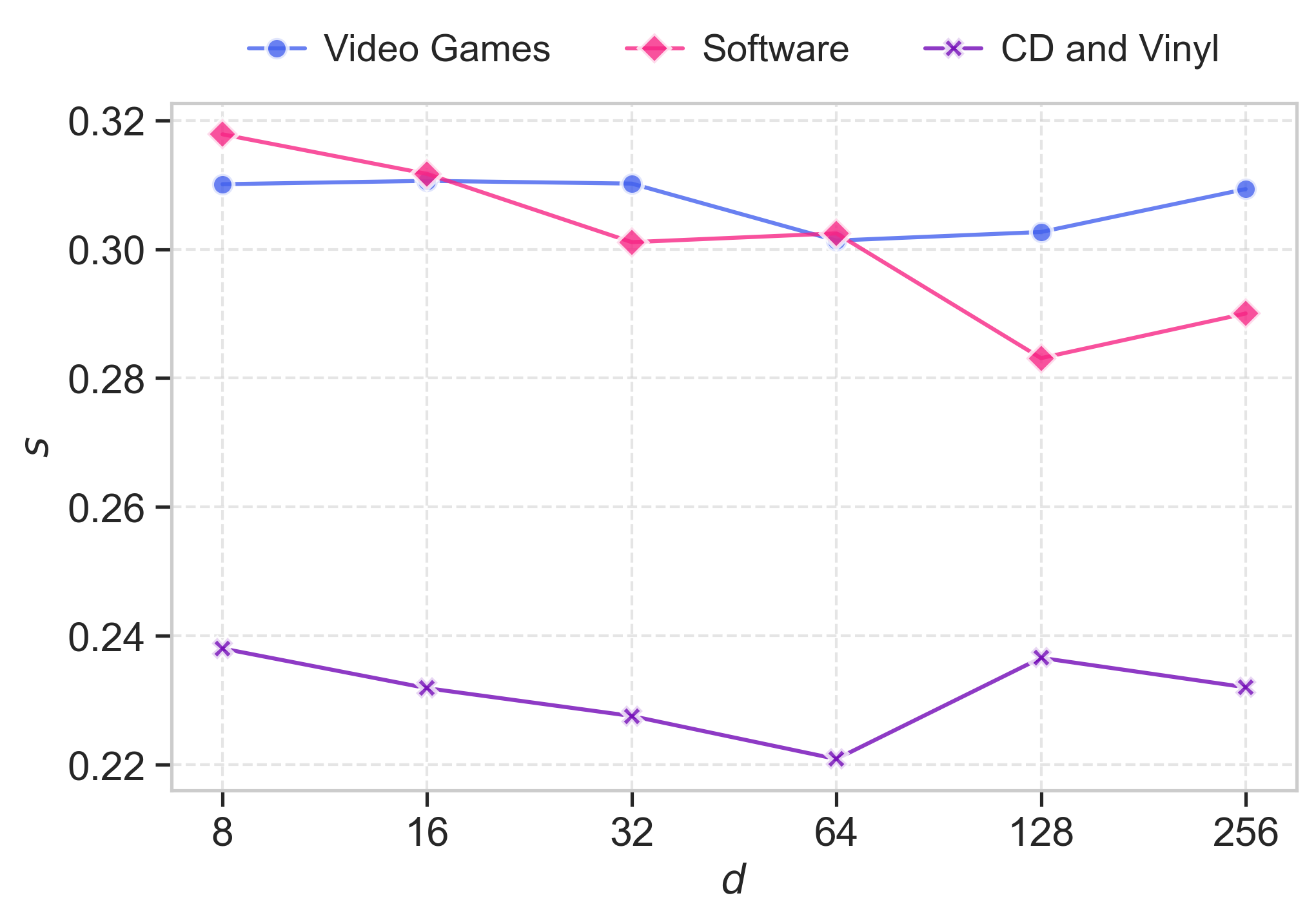}        
    \caption{Influence of latent space dimension, $d$, on DReX.}
    \label{fig:final_d_inf}
\end{figure}

\begin{figure}[b]
    \centering
        \centering
        \includegraphics[height = 5 cm]{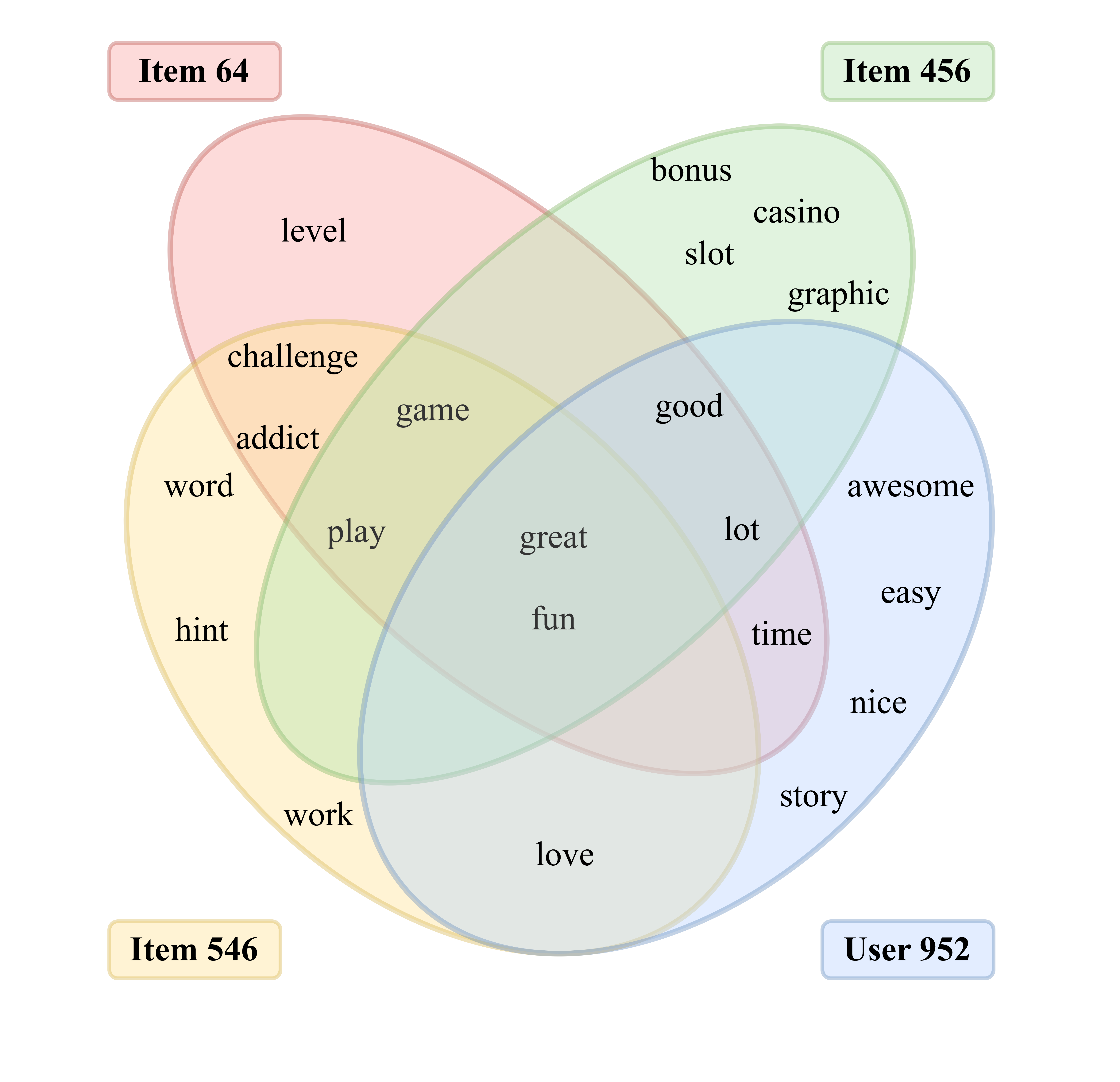}
        
    \caption{Keyword overlap for a random user and its top-3 items.}
    \label{fig:explaination_profile}
\end{figure}

We analyze the impact of latent space dimension, $d$, on our proposed models. Figure \ref{fig:final_d_inf} shows that the \textit{DReX} approach performs optimally at $d=128$ for the Software dataset and at $d=64$ for the other datasets, demonstrating that even with a smaller $d$, the model can effectively capture essential user-item relationships while maintaining model efficiency. A similar trend is observed for \textit{DReX-MLP}.

\begin{table}[!ht]
\renewcommand{\arraystretch}{1}
\centering
\caption{Summary of the Friedman statistical analysis.}
\adjustbox{max width=0.8\linewidth}{
\begin{tabular}{llccccl}
\hline
\textbf{Metric}   & \textbf{$F_F$} & \textbf{Critical Value ($\alpha = 0.05$)} & \textbf{$q_\alpha$} & \multicolumn{1}{l}{\textbf{$\mathcal{N}$}} & \multicolumn{1}{l}{\textbf{$\mathcal{K}$}} & \textbf{CD} \\
\hline 
\textbf{NDCG@1}  & 29.5          & \multirow{3}{*}{2.9961}                                                    & \multirow{3}{*}{2.949}                                & \multirow{3}{*}{3}             & \multirow{3}{*}{7}             & \multirow{3}{*}{5.202}       \\
\textbf{NDCG@5}  & 61        &   &  &  &  &\\
\textbf{NDCG@10} & 124       &   &  &  &  & \\
\hline
\textbf{F1@1}    & 124       & \multirow{3}{*}{2.9961}                                                    & \multirow{3}{*}{2.949}                                & \multirow{3}{*}{3}             & \multirow{3}{*}{7}             & \multirow{3}{*}{5.202}       \\
\textbf{F1@5}    & 75.538  &   &  &  &  &\\
\textbf{F1@10}   & 110       &   &  &  &  & \\
\hline
\end{tabular}
}
\label{tab:friedstat}
\end{table}

To showcase our model's ability to enhance recommendation explainability, we randomly selected a user and their top recommended items based on predicted ratings from the software dataset. Figure \ref{fig:explaination_profile} illustrates the overlap in user and top-3 recommended item keyword profiles, where each profile contains 10 keywords. The shared words highlight the common interests between the user and the items, offering insights into why these items received high predicted ratings.

\begin{figure*}[t]
    \centering
        
        \begin{subfigure}[b]{0.32\textwidth}
            \centering
            \includegraphics[width=0.95\textwidth, trim ={30 40 30 40}, clip]{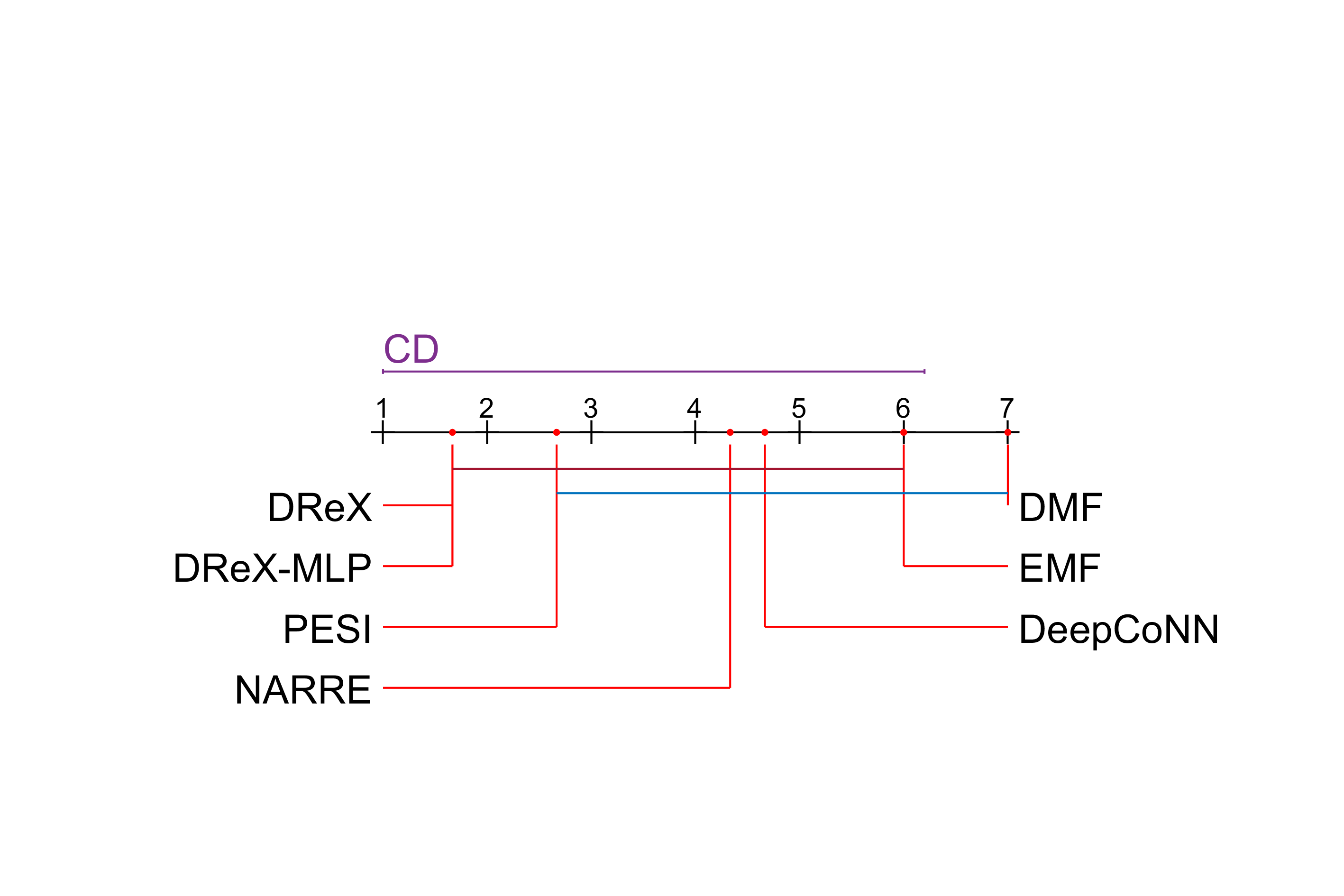}
            \caption{$NDCG@1$}
            \label{fig:n@1_nem}
        \end{subfigure}
        \hfill
        \begin{subfigure}[b]{0.32\textwidth}
            \centering
            \includegraphics[width=0.9\textwidth, trim ={30 40 30 40}, clip]{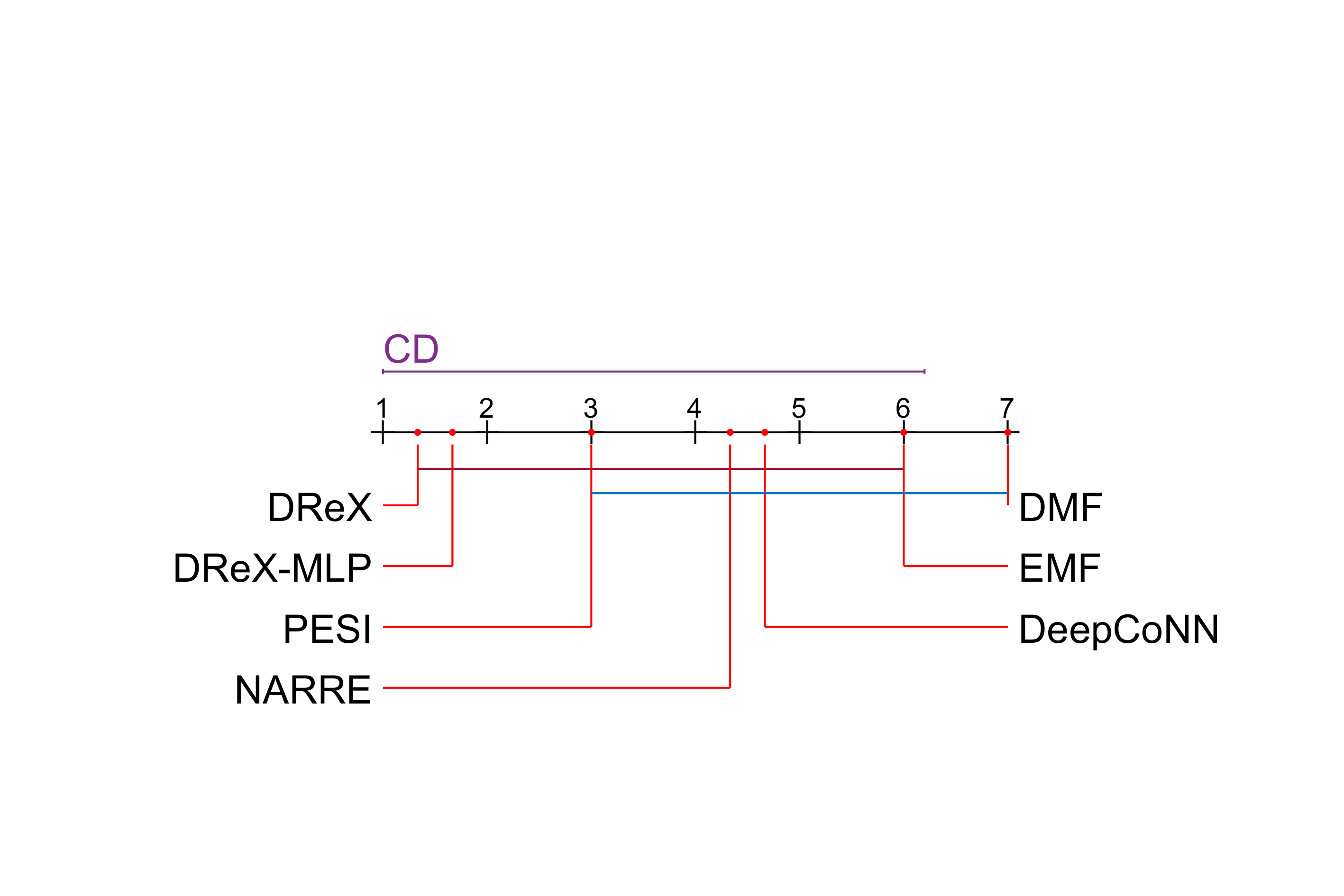}
            \caption{$NDCG@5$}
            \label{fig:n@5_nem}
        \end{subfigure}
        \hfill
        \begin{subfigure}[b]{0.32\textwidth}
            \centering
            \includegraphics[width=0.9\textwidth, trim ={30 40 30 40}, clip]{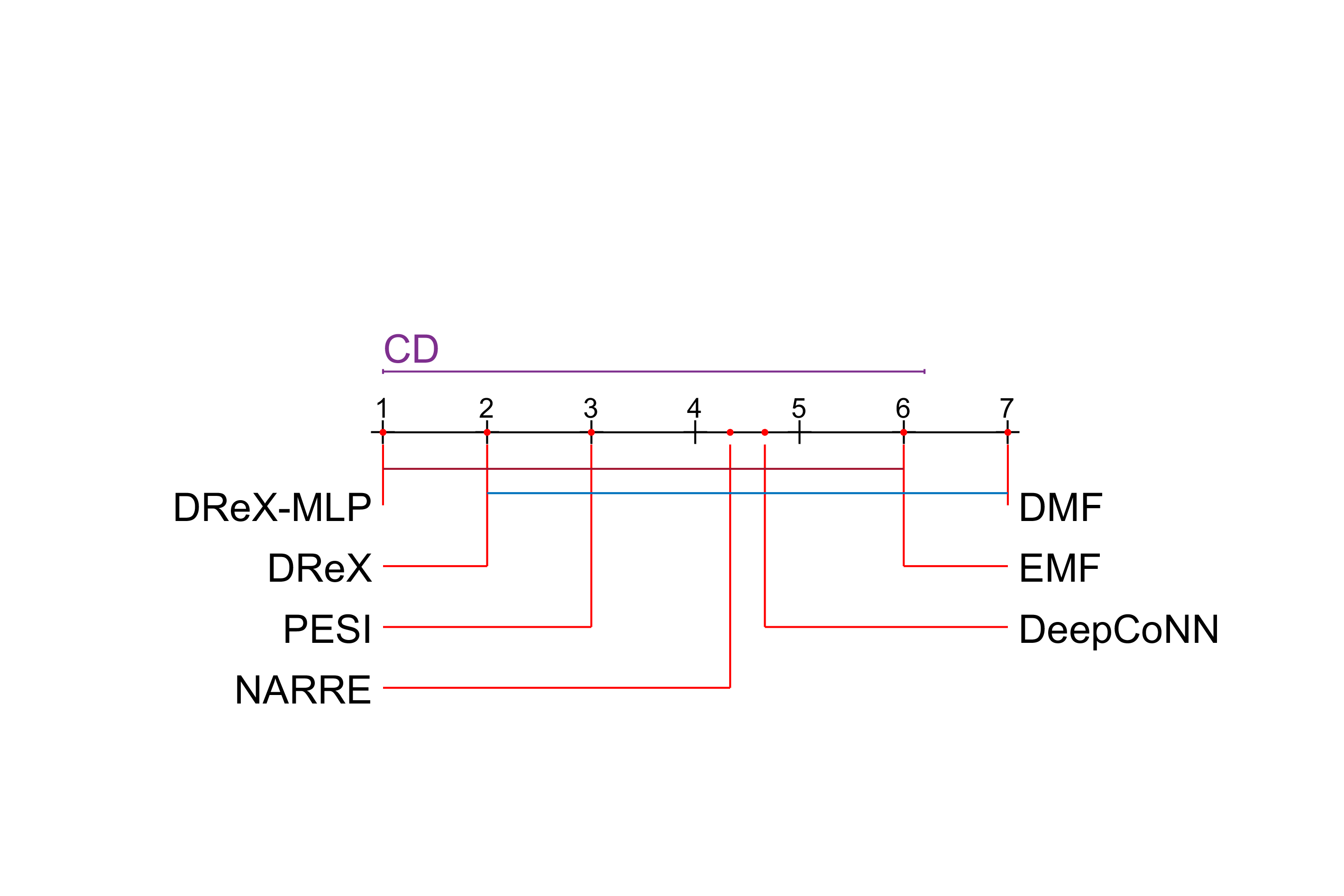}
            \caption{$NDCG@10$}
            \label{fig:n@10_nem}
        \end{subfigure}
        
        \begin{subfigure}[b]{0.32\textwidth}
            \centering
            \includegraphics[width=0.9\textwidth, trim ={30 40 30 40}, clip]{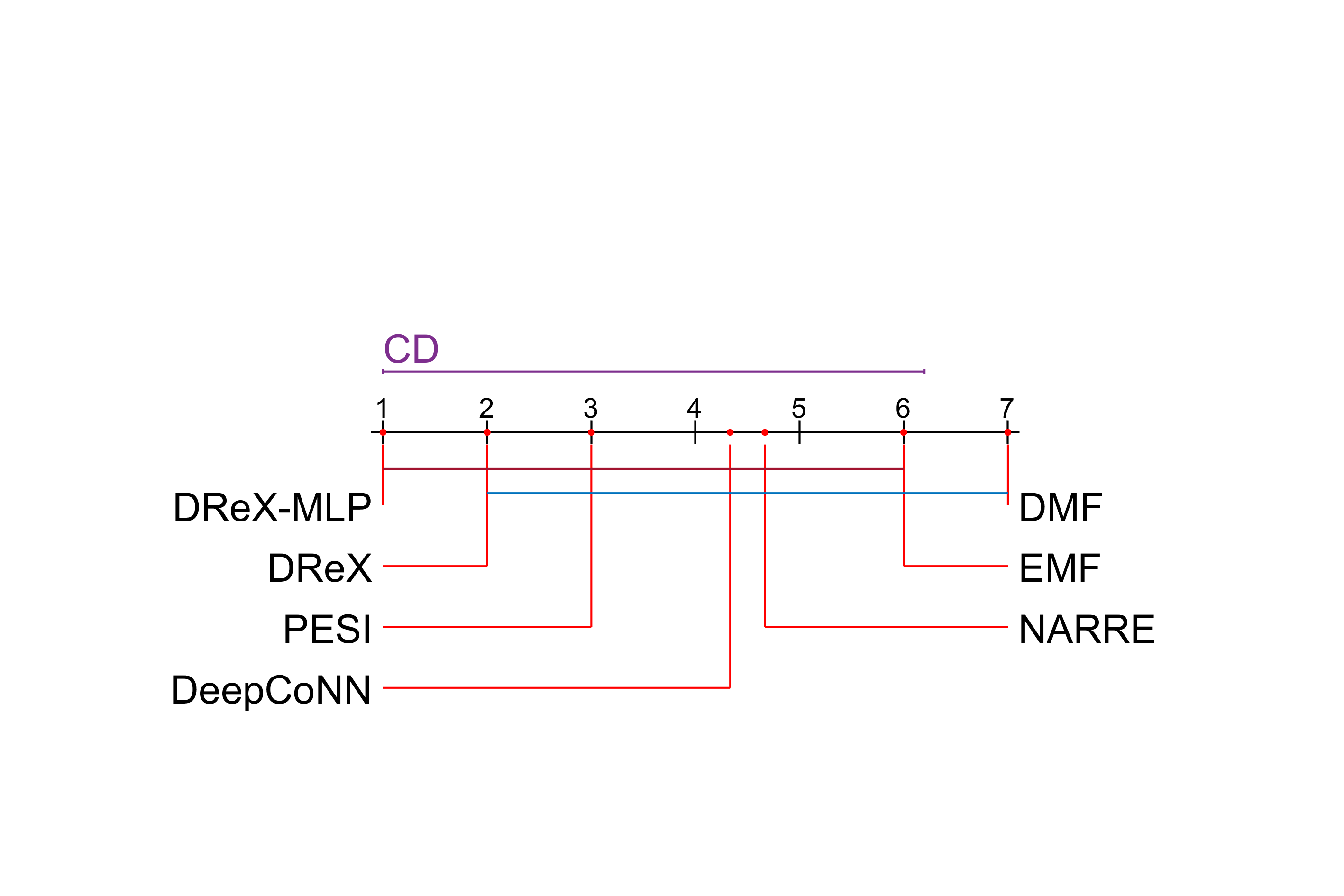}
            \caption{$F_1@1$}
            \label{fig:f1@1_nem}
        \end{subfigure}
        \hfill
        \begin{subfigure}[b]{0.32\textwidth}
            \centering
            \includegraphics[width=0.9\textwidth, trim ={30 40 30 40}, clip]{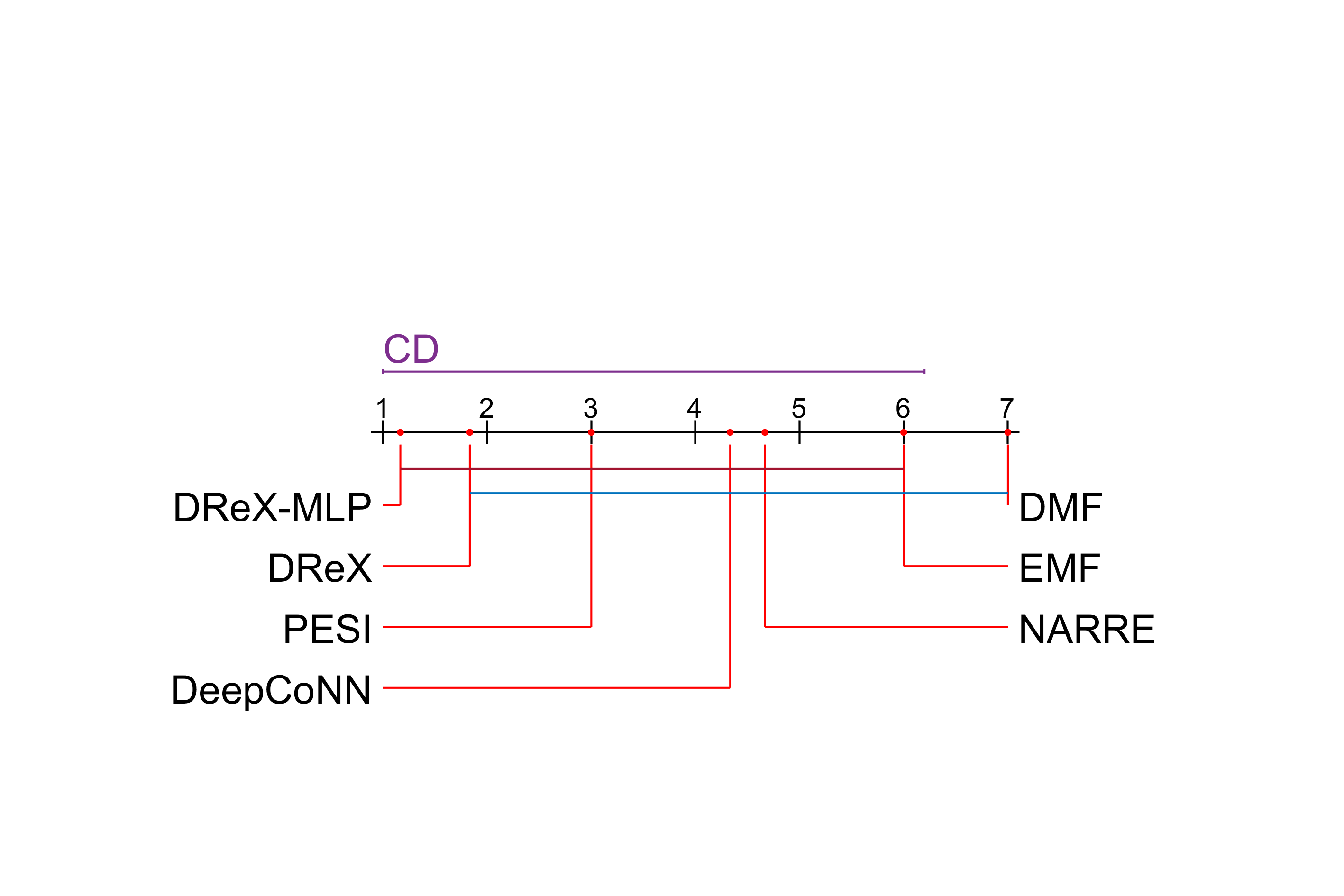}
            \caption{$F1@5$}
            \label{fig:f1@5_nem}
        \end{subfigure}
        \hfill
        \begin{subfigure}[b]{0.32\textwidth}
            \centering
            \includegraphics[width=0.9\textwidth, trim ={30 40 30 40}, clip]{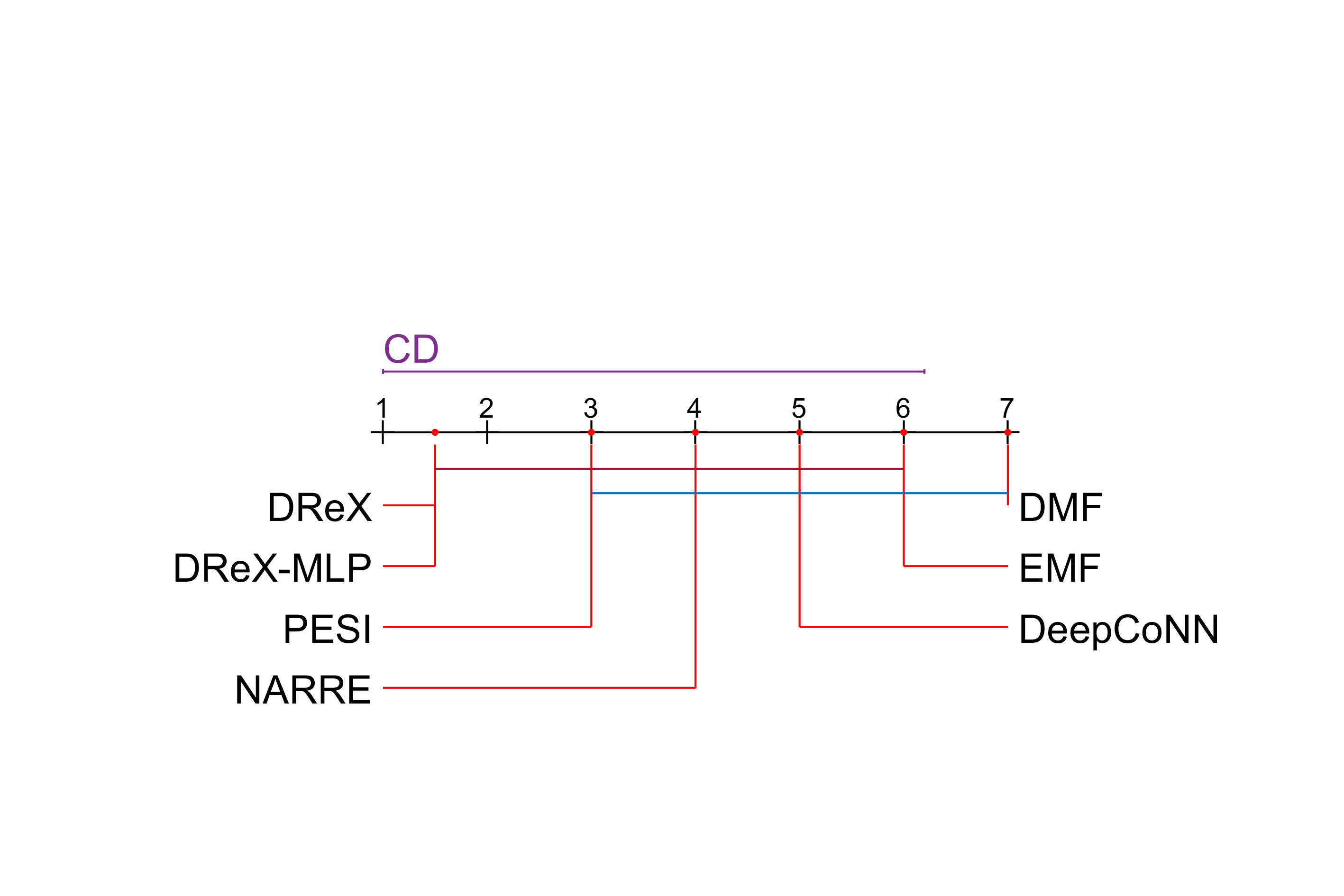}
            \caption{$F_1@10$}
            \label{fig:f1@10_nem}
        \end{subfigure}

    \caption{CD diagrams for comparing algorithms on $NDCG@k$ (a-c) and $F1@k$ (d-f) for $k \in \{1,5,10\}$.}
    \label{fig:criticalDifference}
\end{figure*}

To further validate our experimental results, we conducted a statistical analysis to assess the significance of the proposed approach in terms of its performance over other algorithms. We employed \textit{Friedman test}, which is effective for comparing more than two algorithms over multiple datasets~\cite{demvsar2006statistical}. We present the summary of Friedman analysis for $NDCG@k$ and $F1@k$ for $k = \{1, 5, 10\}$ in Table \ref{tab:friedstat}. As shown in the table, the test rejects the null hypothesis at the significance level of $\alpha = 0.05$, demonstrating a significant difference in the performance of the compared algorithms over the considered evaluation metrics. We then employed yet another statistical test, the \textit{Nemenyi post-hoc test}~\cite{demvsar2006statistical}, which compares the average rank of two algorithms to check if the performances of these algorithms are significantly different. For $\mathcal{K}$ number of comparing algorithms evaluated over $\mathcal{N}$ datasets, the test calculates a critical distance metric $CD$ as $q_\alpha \sqrt{\frac{\mathcal{K}(\mathcal{K}+1)}{6\mathcal{N}}}$. Here,  $q_\alpha$ denotes the critical value. If the average rank difference between two algorithms exceeds the $CD$ value, they are considered significantly different. Figure \ref{fig:criticalDifference} presents the CD diagrams for $NDCG@k$ and $F1@k$ scores for $k \in \{1,5,10\}$. The diagrams show that the proposed framework consistently ranks highest for each metric, confirming its superiority. Similar results were observed for all other values of $k$.

In conclusion, these results highlight the effectiveness of our proposed method, which captures both local interaction-level nuances and global user-item relationships. The local interaction-level representations allow the model to focus on fine-grained, context-specific details, while the fusion into a global representation ensures that broader user-item dynamics are also considered. By integrating these two aspects, our approach balances the local relevance and global context, leading to superior performance.

\section{Conclusion}
\label{concFuture}
In this work, we have introduced \textit{DReX}, a novel explainable multimodal recommendation framework that dynamically learns the user and item representations from multiple modalities. \textit{DReX} uses gated recurrent units to refine global representations by incorporating fine interaction-level features. Unlike traditional approaches, \textit{DReX} jointly processes user and item features to improve the alignment between the learned representations of both entities.  The proposed method is also robust to varying and missing modalities at the interaction level. Additionally, by integrating review text, it maintains user and item keyword profiles to provide recommendation explainability. Extensive experiments across three real-world datasets validate the efficacy of the proposed approach over established baseline methods.
In the future, we plan to explore more sophisticated fusion techniques to better handle capture dependencies across modalities. Generating natural language-based explanations is another promising direction to enhance interpretability. Additionally, integrating richer modalities like images or audio, and extending the proposed framework to cross-domain scenarios could further improve its adaptability and effectiveness in diverse recommendation settings.

\bibliographystyle{unsrt}
\bibliography{ref}

\end{document}